\documentclass[fleqn,usenatbib]{mnras}
\usepackage{newtxtext,newtxmath}
\usepackage[T1]{fontenc}
\usepackage{ae,aecompl}
\usepackage{here}
\usepackage{float}

\usepackage[dvipdfmx]{graphicx}	% Including figure files
\usepackage{amsmath}	% Advanced maths commands
\usepackage{lscape}
\usepackage{graphicx}
\usepackage{multirow}
\usepackage{color}
\usepackage{bm}
\usepackage{comment}
\usepackage{ulem}

\newcommand{\Msun}{\mathrm{M}_{\odot}} 
\newcommand{\Zsun}{\rm{Z}_{\odot}}

\newcommand{\Secref}[1]{Section\:\ref{#1}}
\newcommand{\Appref}[1]{Appendix\:\ref{#1}}
\newcommand{\Tabref}[1]{Table\:\ref{#1}}
\newcommand{\Eqref}[1]{Equation\:(\ref{#1})}

\newcommand{\Figref}[1]{Figure\:\ref{#1}}
\newcommand{\Figsref}[2]{Figures\:\ref{#1}~and~\ref{#2}}

\title[Disc fragmentation across all metallicities]{Protostellar-disc fragmentation across all metallicities}

\author[R. Matsukoba et al.]
{Ryoki Matsukoba,$^{1,2}$\thanks{E-mail: r.matsukoba@tap.scphys.kyoto-u.ac.jp} 
Kei E. I. Tanaka,$^{3,4}$
Kazuyuki Omukai,$^{1}$
Eduard I. Vorobyov$^{5}$ and
\newauthor
Takashi Hosokawa$^{2}$ 
\\
$^{1}$Astronomical Institute, Graduate School of Science, Tohoku University, Aoba, Sendai 980-8578, Japan\\
$^{2}$Department of Physics, Graduate School of Science, Kyoto University, Sakyo, Kyoto 606-8502, Japan\\
$^{3}$Center for Astrophysics and Space Astronomy, University of Colorado Boulder, Boulder, CO 80309, USA\\
$^{4}$National Astronomical Observatory of Japan, National Institutes of Natural Sciences, 2-21-1 Osawa, Mitaka, Tokyo 181-8588, Japan\\
$^{5}$Department of Astrophysics, University of Vienna, Vienna, 1180, Austria
}

\date{Accepted XXX. Received YYY; in original form ZZZ}

\pubyear{2022}

\begin{document}
\label{firstpage}
\pagerange{\pageref{firstpage}--\pageref{lastpage}}
\maketitle

%%% word limit: 250 %%%
\begin{abstract}
Cosmic metallicity evolution possibly creates the diversity of star formation modes at different epochs. Gravitational fragmentation of circumstellar discs provides an important formation channel of multiple star systems, including close binaries. We here study the nature of disc fragmentation, systematically performing a suite of two-dimensional radiation-hydrodynamic simulations, in a broad range of metallicities, from the primordial to the solar values. In particular, we follow relatively long-term disc evolution over 15 kyr after the disc formation, incorporating the effect of heating by the protostellar irradiation. Our results show that the disc fragmentation occurs at all metallicities $1$--$0$\:$\Zsun$, yielding self-gravitating clumps. Physical properties of the clumps, such as their number and mass distributions, change with the metallicity due to different gas thermal evolution. For instance, the number of clumps is the largest for the intermediate metallicity range of $10^{-2}$--$10^{-5}$\:$\Zsun$, where the dust cooling is effective exclusively in a dense part of the disc and causes the fragmentation of spiral arms. The disc fragmentation is more modest for $1$--$0.1$\:$\Zsun$ thanks to the disc stabilization by the stellar irradiation. Such metallicity dependence agrees with the observed trend that the close binary fraction increases with decreasing metallicity in the range of $1$--$10^{-3}$\:$\Zsun$. 
\end{abstract}


%%%%%%%%%%%%%%%%%%%%%%%%%%%%%%%%%%%%%%%%%%%%%%%%%%
%%%%%%%%%%%%%%%%% BODY OF PAPER %%%%%%%%%%%%%%%%%%

%%%%%%%%%%%%%%%%%%%%%%%%%%%%%%%%%%%%%%%%%%%%
%%%%%%%%%%%%%%%%%%%%%%%%%%%%%%%%%%%%%%%%%%%%
%%% SECTION 1 %%%%
\section{Introduction}
\label{Sec:1}
Stars in the contemporary and early universe differ in the mass. 
Those in the local universe at the solar metallicity have a peak in the initial mass function at $\lesssim 1$\:$\Msun$ \citep{Kroupa:2002, Chabrier:2003, Bastian:2010}, while first stars forming from the primordial gas (or Pop III stars) have much higher masses of $\sim 10$--$100$\:$\Msun$ according to numerical simulations \citep[e.g.][]{Hirano:2014, Hirano:2015,Susa:2014}. 
This indicates that the typical stellar mass has been changing over cosmic time along with the accumulation of metals in a star-forming gas. 
Some metallicity-dependent star-formation processes must have caused this transition. 

%----------------------------------------------------------------%

One such process is gravitational fragmentation of a prestellar cloud.  
Theories tell that just a little bit of metals and dust grains in a star-forming gas make a big difference in this regard.
Semi-analytical models predict that those clouds are vulnerable to fragmentation provided the metallicity is $\gtrsim 10^{-5}$\:$\Zsun$ owing to the dust cooling operating at densities $>10^{10}$\:${\rm cm}^{-3}$ \citep{Omukai:2000, Schneider:2003, Schneider:2006, Omukai:2005, Omukai:2010, Chiaki:2014, Chiaki:2015}.
Numerical simulations demonstrate that the cloud fragmentation indeed occurs, forming low-mass clumps with $\sim 0.01$\:$\Msun$, unlike in the case of  $0$\:$\Zsun$ \citep{Clark:2008, Dopcke:2011, Dopcke:2013, Meece:2014, Smith:2015, Chiaki:2016, Safranek-Shrader:2016}. 
This enables the formation of low-mass ($\lesssim 1$\:$\Msun$) stars in the early universe.

%-------------------------------------------------------------------%

Fragmentation takes place also in a later stage of star formation after the birth of a protostar; in a circumstellar disc \citep[][]{Kratter:2016}. 
A disc accreting gas from the surrounding envelope tends to fragment by gravitational instability, producing self-gravitating clumps. 
Some numerical simulations of Pop III star formation report the formation of low-mass clumps with $\sim 0.1$--$1$\:$\Msun$ as a result of the disc fragmentation \citep[e.g.][]{Machida:2008, Stacy:2010, Clark:2011, Greif:2012, Machida:2013, Vorobyov:2013, Hosokawa:2016, Stacy:2016, Hirano:2017-9a, Riaz:2018, Susa:2019, Sharda:2020, Sugimura:2020, Wollenberg:2020, Kimura:2021, Prole:2022}. 
The disc fragmentation is also supposed to provide 
a channel for binary and planet formation in the present-day star formation \citep[e.g.][and references therein]{Zhao:2020}. 
Understanding metallicity dependence of the disc fragmentation is thus a clue to bridge the gap between star formation in the early and contemporary universe.
\cite{Tanaka:2014} studied the metallicity dependence of the disc fragmentation using a one-dimensional disc model assuming steady accretion. 
They predicted that discs are most vulnerable to gravitational fragmentation in a metallicity range of $10^{-3}$--$10^{-5}$\:$\Zsun$ due to efficient dust cooling. 

%--------------------------------------------------------------------%

Observations also support the idea that low metallicity discs are more prone to fragment. \cite{Moe:2019} reported that the close binary fraction (separation less than 10\:au) of solar-type stars inversely correlates with the metallicity at $-3<[{\rm Fe/H}]<0$. 
Although difficult to statistically examine the binary fraction at lower metallicities, \citet{Schlaufman:2018} reported the discovery of a close binary with ${\rm [Fe/H]}=-4.07$. The disc fragmentation is a plausible mechanism to form such close binary systems.

%---------------------------------------------------------------------%

Self-gravitational hydrodynamics simulations are powerful tools to study metallicity dependence of the disc fragmentation, and have demonstrated that the disc fragmentation is intense at metallicities $10^{-3}$--$10^{-5}$\:$\Zsun$ \citep{Machida:2015, Chon:2021-12, Shima:2021, Chiaki:2022}, as predicted by \cite{Tanaka:2014}. 
Those previous studies, however, still have the following limitations. 
For example, the metallicity range considered differs in different studies; $1$--$10^{-2}$\:$\Zsun$ in \cite{Vorobyov:2020-9} and $10^{-3}$--$0$\:$\Zsun$ in \cite{Shima:2021} and \cite{Chiaki:2022}. 
The limited range of metallicity studied hinders thorough understanding of the difference in the star formation in the early and contemporary universe.
Although \cite{Machida:2015}'s calculation covered a wider metallicity range of $1$--$0$\:$\Zsun$, they followed the evolution only up to $\sim 100$\:yr after the protostar formation, far shorter than the entire duration of accretion, $\sim 10^{5}$\:yr. \cite{Chon:2021-12} followed longer-term ($\sim 10^5$\:yr) evolution in a metallicity range of $0.1$--$10^{-6}$\:$\Zsun$, without incorporating the disc irradiation heating from the accreting protostars, which effectively stabilizes the disc, as known for the solar-metallicity case \citep{Matzner:2005, Vorobyov:2010}.

%----------------------------------------------%

In this work, we systematically study the metallicity dependence of the disc fragmentation beyond the limitations of the previous studies. 
We consider the cases with different metallicities of the full coverage, $1$--$0$\:$\Zsun$. 
We follow the long-term evolution for 15\:kyr after the disc formation, using two-dimensional radiation-hydrodynamic simulations, where
we consistently solve the gas thermal evolution affected by the stellar irradiation heating. 
To see the metallicity dependence quantitatively, we analyze physical properties of self-gravitating clumps produced by disc fragmentation, such as their number and mass distributions for each case.

%---------------------------------------------------------------------------%

The structure of the rest of the paper is as follows. We describe our simulation method in \Secref{Sec:2}. We present the basic results in \Secref{Sec:3}, where we clarify metallicity dependence of the disc fragmentation. We provide discussion and summary in \Secref{Sec:4} and \Secref{Sec:5}, respectively.

%%%%%%%%%%%%%%%%%%%%%%%%%%%%%%%%%%%%%%%%%%%%
%%%%%%%%%%%%%%%%%%%%%%%%%%%%%%%%%%%%%%%%%%%%
%%% SECTION 2 %%%%
\section{Method}
\label{Sec:2}

We perform a suite of two-dimensional simulations integrating the vertical structure of a disc. In what follows, we briefly describe the method for the self-gravitational radiation-hydrodynamic simulations and our initial conditions. Further details are also provided in \cite{Vorobyov:2020-6, Vorobyov:2020-9}.

%%%%%%%%%%%%%%%%%%%%%%%%%%%%%%%%%%%%%%%%%%%%
%%% SECTION 2-1 %%%%
\subsection{Two-dimensional self-gravitational radiation-hydrodynamic simulations}
\label{Sec:2-1}

We use polar-coordinate $(r,\:\phi)$ grids with $512 \times 512$ spatial zones as a default setting. The grids are logarithmically spaced in the radial direction and have equal spacing in the azimuthal direction. A sink cell with the radius of 5\:au is introduced at the centre of the computational domain. The position of the outer boundary varies according to different models as described in \Secref{Sec:2-2}.

%-----------------------------------------------------------------------%

To follow the time evolution of the gas surface density, velocity, and temperature, 
we solve the vertically-integrated equations of continuity, motion, and energy transport: 
\begin{align}
&\frac{\partial\Sigma}{\partial t} = -\nabla\cdot\left( \Sigma\bm{u} \right), \label{Eq:1} \\
&\frac{\partial}{\partial t}\left( \Sigma\bm{u} \right) 
+ \nabla\cdot \left( \Sigma\bm{u}\otimes\bm{u} \right)
= -\nabla P + \Sigma\bm{g} + \nabla\cdot\bm{\Pi}, \label{Eq:2} \\
&\frac{\partial e}{\partial t} + \nabla\cdot\left( e\bm{u} \right)
= -P\left( \nabla\cdot\bm{u} \right) - Q_{\mathrm{tot}} 
+ \left( \nabla \bm{u} \right):\bm{\Pi}, \label{Eq:3}
\end{align}
where $\Sigma$ is the gas surface density, 
$\bm{u}=u_{r}\bm{\hat{r}}+u_{\phi}\bm{\hat{\phi}}$ is the planar velocity, 
$P$ is the vertically-integrated gas pressure, 
$\nabla=\bm{\hat{r}}\partial/\partial r+\bm{\hat{\phi}}r^{-1}\partial/\partial\phi$ is the derivative operator, 
$\bm{g}=g_{r}\bm{\hat{r}}+g_{\phi}\bm{\hat{\phi}}$ is the gravitational acceleration, 
$e$ is the internal energy per unit area,
$Q_{\mathrm{tot}}$ is the total cooling rate per unit area, 
and $\bm{\Pi}$ is the viscous stress tensor. 

%-----------------------------------------------------------------------%

The gas pressure and internal energy are related by the ideal-gas equation of state, 
\begin{align}
P = (\gamma-1)e,
\label{Eq:4}
\end{align}
with the adiabatic exponent $\gamma$. 
The viscous stress tensor is 
\begin{align}
\bm{\Pi} = 2\Sigma\nu\left[ \nabla\bm{u}-\frac{1}{3}(\nabla\cdot\bm{u})\bm{\mathrm{e}} \right],
\label{Eq:9}
\end{align}
where $\bm{\mathrm{e}}$ is the unit tensor and 
$\nu$ is the kinematic viscosity, which is given according to the $\alpha$ description \citep{Shakura:1973}.
We set $\alpha=10^{-4}$ uniformly in the space to account for residual turbulence in the disc midplane owing possibly to the magnetorotational instability at the surface layers of the disc.
We note that the gravitational torque, rather than the viscosity,  dominates the angular momentum transfer in our simulations, as is the case for solar-metallicity protostellar discs in their early stages of evolution \citep{Vorobyov:2009-3}.
The self-gravitational potential created by the gas is 
\begin{align}
\Phi_{\mathrm{gas}}(r,\:\phi) &= -G\int^{r_{\mathrm{out}}}_{r_{\mathrm{sc}}} r'\mathrm{d}r' \notag\\
&\times \int^{2\pi}_{0} \frac{\Sigma(r',\:\phi')}{\sqrt{r'^{2}+r^{2}-2rr'\mathrm{cos}(\phi'-\phi)}}~\mathrm{d}\phi',
\label{Eq:8}
\end{align}
where $G$ is the gravitational constant, and $r_{\mathrm{sc}}$ is the sink radius of 5\:au. The gravitational acceleration by the gas potential is 
\begin{align}
\bm{g}_{\mathrm{gas}} = -\nabla\Phi_{\mathrm{gas}}(r,\:\phi). 
\end{align}
We assume that the sink cell represents a single central star accreting the gas.  
The gravitational acceleration by the central star with the mass $M_{\ast}$ is 
\begin{align}
\bm{g}_{\mathrm{star}} = -\frac{GM_{\ast}}{r^{2}}\bm{\hat{r}}. 
\end{align}
The sum of the two components gives the gravitational acceleration, i.e.
$\bm{g}=\bm{g}_{\mathrm{gas}}+\bm{g}_{\mathrm{star}}$. We assume hydrostatic equilibrium in the direction perpendicular to the $r$--$\phi$ plane everywhere \citep{Vorobyov:2009-3}:
\begin{align}
\rho c_{\mathrm{s}}^2 = \frac{\pi}{2}G\Sigma^2 + \frac{2GM_{\mathrm{star}}\rho}{r} 
\left[ 1-\left( 1+\frac{\Sigma}{2\rho r} \right)^{-1/2} \right],
\end{align}
where $\rho$ is the mass density, and $c_{\mathrm{s}}$ is the sound speed defined as 
\begin{align}
c_{\mathrm{s}} = \sqrt{\gamma\frac{k_{\mathrm{B}}T}{\mu m_{\mathrm{H}}}},
\label{Eq:cs}
\end{align}
where $k_{\mathrm{B}}$ is the Boltzmann constant, $T$ is the gas temperature, $\mu$ is the mean molecular weight, and $m_{\mathrm{H}}$ is the mass of a hydrogen nucleus.

%-------------------------------------------------%

The total cooling rate is the sum of the contribution by each process as follows \citep{Vorobyov:2020-6}: 
\begin{align}
Q_{\mathrm{tot}} = Q_{\mathrm{cont,gas}} + Q_{\mathrm{cont,dust}} + Q_{\mathrm{mole}} + 
Q_{\mathrm{metal}} + Q_{\mathrm{chem}}, 
\end{align}
where $Q_{\mathrm{cont,gas}}$ and $Q_{\mathrm{cont,dust}}$ are the rates by continuum emission of gas and dust, $Q_{\mathrm{mole}}$ by line emission of H$_2$ and HD molecules, $Q_{\mathrm{metal}}$ by fine-structure line emission of O~{\sc I} ($63~\mu$m) and C~{\sc II} ($158~\mu$m), and $Q_{\mathrm{chem}}$ is the chemical cooling rate. 
The dust continuum term $Q_{\mathrm{cont,dust}}$ represents the process by which the energy of the gas is transferred to the dust grains by collision and released into the radiation field. This process acts as cooling for the gas 
when the gas temperature is higher than the dust temperature, while as heating otherwise. 
We evaluate the dust temperature $T_{\rm d}$ by solving the energy balance between the photon absorption and emission at the dust surface and collisional energy exchange with the gas \citep{Omukai:2010,Vorobyov:2020-6}: 
\begin{align}
\kappa_{\rm P,d}B(T_{\rm d}) = \kappa_{\rm P,d}J + \Gamma_{\rm coll},
\label{Eq:thermal_balance}
\end{align}
where $\kappa_{\rm P,d}$ is the Planck mean opacity for dust grains, $B(T_{\rm d}) = \sigma_{\rm SB}T_{\rm d}^4/\pi$ is the intensity for the black body at $T_{\rm d}$ with the Stefan-Boltzmann constant $\sigma_{\rm SB}$, $J$ is the mean intensity, which includes the effects of the central stellar irradiation and uniform background radiation, and $\Gamma_{\rm coll}$ is the heating rate of dust through collisions with gas particles \citep{Hollenbach:1979}. Note the collisional heating for dust grains is cooling for the gas. We calculate the Planck mean opacity  $\kappa_{\rm P,d}$ using the table given in \cite{Semenov:2003}.
The chemical cooling term can also be negative, in this case it works as heating. 
The terms $Q_{\mathrm{cont,dust}}$ and $Q_{\mathrm{metal}}$ are proportional to the metallicity in the optically thin regime.
We note that the photon trapping effect for these cooling terms is also considered by evaluating the optical depth through the disc as in \cite{Vorobyov:2020-6}.

%-------------------------------------------------------------------%

In self-gravitational hydrodynamic simulations, artificial fragmentation can be triggered 
if the Jeans length $l_{\mathrm{J}}$ is only poorly resolved with less than 4 grid cells:
\begin{align}
4 x_{\mathrm{grid}} > l_{\mathrm{J}} =  \sqrt{\frac{\pi c_{\mathrm{s}}^2}{G\rho}},
\label{Eq:Truelove}
\end{align}
where $x_{\mathrm{grid}}$ is the grid size \citep{Truelove:1997}. Subsequent studies also suggest that even higher resolution is necessary to capture the gravity-driven turbulence \cite[e.g.][]{Federrath:2011,Lichtenberg:2015}.
To prevent the artificial fragmentation, we suppress cooling by multiplying the total rate $Q_{\mathrm{tot}}$ by a function:
\begin{align}
C_{\mathrm{limit}} = \left\{ 
\begin{array}{ll}
\exp \left[-\left(\frac{\xi-1}{0.1}\right)^{2} \right]~~~&(\xi \ge 1), \\
1 &(\xi < 1),
\label{Eq:Climit}
\end{array}
\right.
\end{align}
\begin{align}
\xi = f_{\mathrm{limit}}\frac{x_{\mathrm{grid}}}{l_{\mathrm{J}}},
\end{align}
where $f_{\mathrm{limit}}$ is an arbitrary coefficient \citep{Hosokawa:2016}, which is 6 in this study. The function $C_{\mathrm{limit}}$ becomes smaller than 1 where the Jeans length is shorter than six times the grid size, reducing the cooling rate and preventing artificial fragmentation.

%---------------------------------------------------------------------------------------------------------%

We assume that the central star accretes the gas flowing into the sink cell at each time step. 
We consider the irradiation by the accreting protostar, which contributes to the disc stabilization by increasing the dust and gas temperatures (see Equation \ref{Eq:thermal_balance} and \Secref{Sec:4-1}). 
The protostellar luminosity consists of two components; accretion luminosity and stellar intrinsic luminosity.
We evaluate the intrinsic luminosity from pre-calculated stellar evolutionary tracks.
While \cite{Vorobyov:2020-9} use the tracks obtained by the STELLAR code \citep{Yorke:2008}, we use the tracks calculated as in \cite{Hosokawa:2009} for all the metallicities we consider. Our tracks for $10^{-2},$\:$10^{-3}$, and $0$\:$\Zsun$ are identical to those used in \cite{Fukushima:2020-7}.
The evolutionary tracks provide the stellar radius and intrinsic luminosity as functions of the stellar mass and accretion rate. We use the time-averaged accretion rate for the previous five years because it fluctuates significantly in time. 

%--------------------------------------------------------------------------------------------------%

We also add a uniform background radiation field over the entire computational domain,
which imposes a floor temperature.
The background radiation is assumed to be 10\:K, the temperature of the cosmic microwave background at the redshift of $z=3$. The floor temperature is reached only for the cases of $1$--$10^{-2}$\:$\Zsun$ in our simulations. 

%------------------------------------------------------------------------------------------%

We solve the non-equilibrium chemical network of 8 species H, H$_{2}$, H$^+$, H$^{-}$, D, HD, D$^+$, and e$^-$ with 27 reactions (see \citealt{Vorobyov:2020-6}).
We define the chemical fraction as $y(\mathrm{i})\equiv n(\mathrm{i})/n_{\mathrm{H}}$, 
where $n(\mathrm{i})$ and $n_{\mathrm{H}}$ are the number densities of a species $\mathrm{i}$ and the hydrogen nuclei, respectively. We assume neutrality for helium, and its chemical fraction is $y(\mathrm{He})=8.333\times10^{-2}$. 
The deuterium fraction is $y(\mathrm{D})=3\times10^{-5}$. We adopt the standard values of the metal contents in the solar neighbourhood for the case of $1$\:$\Zsun$; $y(\mathrm{C})=9.27\times10^{-5}$ and $y(\mathrm{O})=3.57\times10^{-4}$. These values both linearly scale with decreasing metallicity. We also assume that all the carbon and oxygen nuclei always exist as C$^+$ ions and O atoms. Regarding dust grains, we use the standard composition and size distributions in the Galactic interstellar medium \citep[][]{Mathis:1977}. 
We make the gas-dust mass ratio vary in proportion to the metallicity.

%%%%%%%%%%%%%%%%%%%%%%%%%%%%%%%%%%%%%%%%%%%%
%%% SECTION 2-2 %%%%
\subsection{Initial conditions and setups}
\label{Sec:2-2}

%%%%% TABLE 1 %%%%%
\begin{table*} 
 \begin{center}
 \caption{The initial properties of the prestellar cloud cores}
 \label{Tab:2_LM}
  \scalebox{1.0}[1.0]{ 
  {\begin{tabular}{c c c c c c c c} 
     \hline \hline
     metallicity & $T_{0}$ & $c_{\mathrm{s,0}}$ & $\Sigma_0$ & $r_0$ & $r_{\rm{out}}$ & $\Omega_0$ & $M_{\rm{cloud}}$ \\ %\hline
     $\Zsun$& K & km s$^{-1}$ & g cm$^{-2}$ & pc & pc & km s$^{-1}$ pc$^{-1}$ & M$_{\odot}$ \\ \hline
     1              & 10  & 0.19 & 0.022 & 0.032 & 0.19 & 0.60 & 3.4 \\
     10$^{-1}$ & 33  & 0.36 & 0.041 & 0.059 & 0.35 & 0.60  & 21 \\
     10$^{-2}$ & 43  & 0.51 & 0.057 & 0.083 & 0.50 & 0.60  & 60 \\
     10$^{-3}$ & 39  & 0.51 & 0.058 & 0.083 & 0.50 & 0.60  & 61 \\
     10$^{-4}$ & 72  & 0.70 & 0.079 & 0.11   & 0.68 & 0.60  & 160 \\
     10$^{-5}$ & 270 & 1.4 & 0.15  & 0.22   & 1.3  & 0.60  & 1200 \\
     10$^{-6}$ & 310 & 1.4 & 0.16  & 0.24   & 1.4  & 0.60  & 1400 \\
     0              & 310 & 1.4 & 0.16   & 0.24   & 1.4  & 0.60  & 1400 \\ \hline
  \end{tabular}}
  }
  \\
 \end{center}
\end{table*}

We start the simulation from a prestellar cloud core. We make the core gravitationally unstable to initiate the collapse. It takes about twice the free-fall time for a circumstellar disc to appear when the centrifugal radius of the accreting gas exceeds the sink radius. We study the degree of the disc fragmentation through the subsequent evolution at a given metallicity.

%------------------------------------------------------%

The initial core profile has a plateau with a uniform surface density around the centre \citep{Vorobyov:2010}: 
\begin{align}
\Sigma = \frac{\Sigma_0}{\sqrt{1+\left( r/r_0 \right)^2}},
\label{Eq:SurfaceDensity_Init}
\end{align}
\begin{align}
\Omega = 2\Omega_0 \left(\frac{r_0}{r}\right)^2\left[ \sqrt{1+\left(\frac{r}{r_0}\right)^2}-1 \right], 
\end{align}
where $\Sigma_{0}$ is the surface density at the plateau, $\Omega_{0}$ is the angular velocity at the plateau, and $r_{0}$ is the plateau radius. The above profile emerges when an axisymmetric cloud gravitationally collapses with constant angular momentum \citep{Basu:1997}. The plateau radius $r_{0}$ is proportional to the Jeans length
\begin{align}
r_{0} = A\frac{c_{\mathrm{s},0}}{\sqrt{\pi G\rho_{0}}},
\label{Eq:r0}
\end{align}
where $A$ is a constant parameter, $c_{\mathrm{s},0}$ is the initial sound speed, 
and $\rho_{0} = 2.2\times10^{-19}{\rm\:g\:cm^{-3}}$ is the initial mass density ($n_{{\rm H},0}=10^{5}{\rm\:cm^{-3}}$ in the number density).
The initial mass and the plateau surface densities are related as $\Sigma_0=r_0 \rho_0$. The constant $A$ is a parameter defining the initial density perturbation. We set $A=\sqrt{1.2}$, with which the ratio of the cloud thermal and gravitational energies is $0.77$ for all the cases. We also set the ratio of the rotational and gravitational energies as 0.01 by adjusting the plateau angular velocity $\Omega_{0}$. 

%------------------------------------------------------%
%%%%% FIGURE 1 %%%%%
\begin{figure}
 \begin{center}
 \begin{tabular}{c} 
  {\includegraphics[width=0.95\columnwidth]{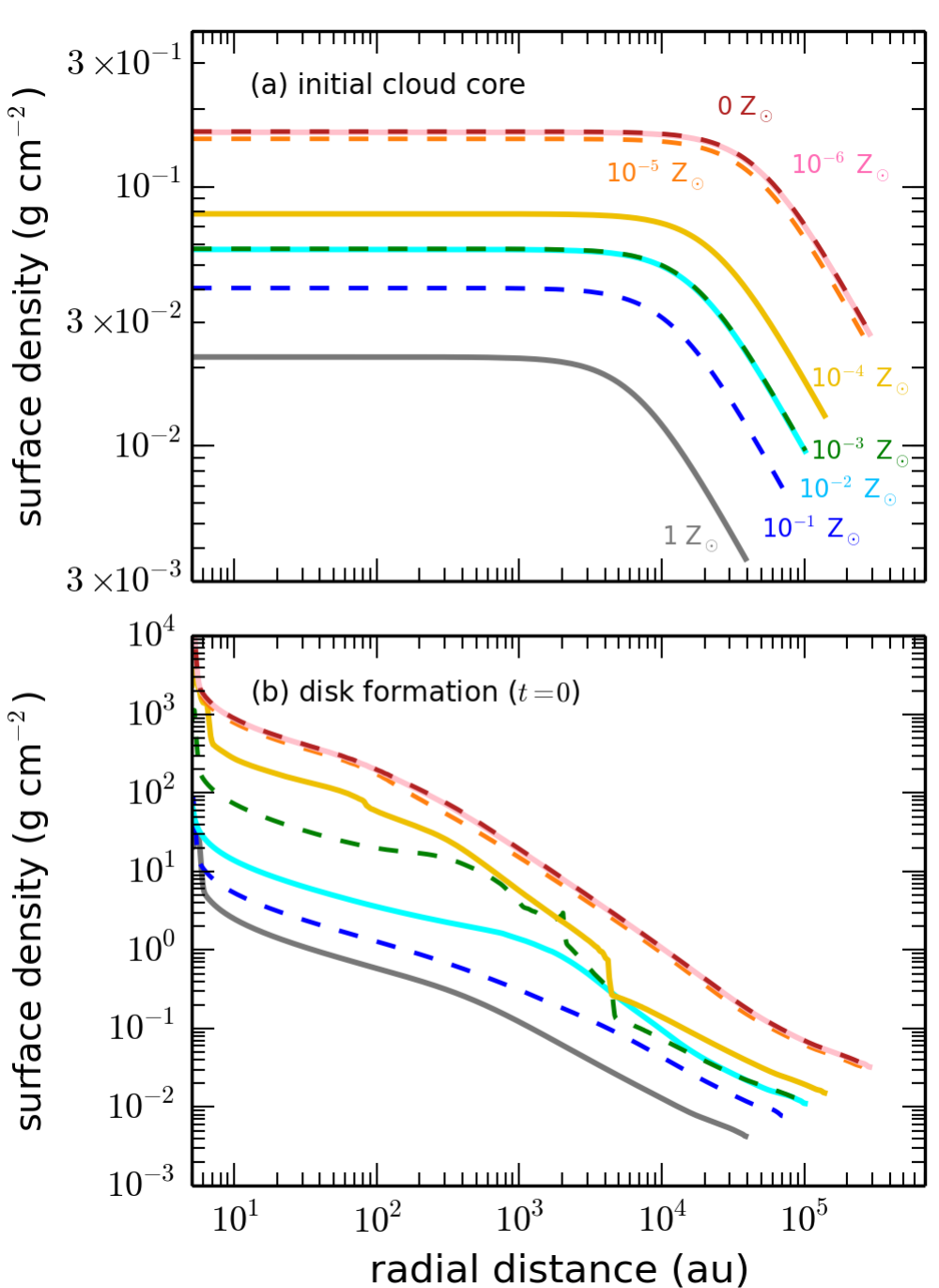}}
 \end{tabular}
 \caption{ 
 Radial profiles of the azimuthally-averaged surface density. 
 The two panels show the snapshots at different epochs: (a) the initial states of the simulations, and (b) the disc formation, which is defined as $t=0$ (see the text). The line colours represent the metallicities of $1$ (black), $10^{-1}$ (blue), $10^{-2}$ (cyan), $10^{-3}$ (green), 
 $10^{-4}$ (yellow), $10^{-5}$ (orange), $10^{-6}$ (pink), and $0$\:$\Zsun$ (red).
 }
 \label{Fig:Surface_Init}
 \end{center}
\end{figure}
%

%------------------------------------------------------%

\Tabref{Tab:2_LM} summarizes the models with different metallicities we consider. To determine the initial gas temperature and chemical fractions, both of which depend on the metallicity, we use a one-zone model developed by \cite{Omukai:2005}. We follow the evolution of the central density, temperature, and chemical fractions of a collapsing core until the central density reaches $n_{\mathrm{H},0}=10^5$\:cm$^{-3}$. We assign the values obtained by the one-zone model homogeneously to the entire computational domain. The outer boundary of the computational domain $r_{\mathrm{out}}$ is set as six times larger than the plateau radius $r_0$, to ensure that the initial configuration is marginally unstable to gravitational collpase. We show the surface density profiles of the prestellar cloud obtained from Equation~(\ref{Eq:SurfaceDensity_Init}) in \Figref{Fig:Surface_Init}a. The plateau surface density $\Sigma_{0}$ is smaller with higher metallicity because the initial sound speed and thus the plateau radius are smaller. We can calculate the total cloud mass $M_{\mathrm{cloud}}$ by using the initial surface density profile, i.e. ranging from the smallest value of 3.4\:$\Msun$ at $1$\:$\Zsun$ to the largest 1400\:$\Msun$ at $0$\:$\Zsun$ (\Tabref{Tab:2_LM}).

%----------------------------------------------------------------------------------------------------%

As mentioned earlier in this section, the initial prestellar cloud gravitationally collapses, and then the circumstellar disc forms once the centrifugal radius of the infalling gas becomes larger than the sink cell radius of $5{\rm\:au}$. \Figref{Fig:Surface_Init}b shows the surface density profiles at the disc formation. We define the epoch of disc formation as $t=0$, and then follow the evolution until $t=15{\rm\:kyr}$. As shown below, the stellar mass exceeds $20$\:$\Msun$ by the end of the simulations in the cases of $10^{-5},~10^{-6}$, and $0$\:$\Zsun$ owing to high accretion rates. We do not incorporate the ionizing feedback from such massive stars, which disturbs the accretion flow and limits the stellar mass growth \citep{McKee:2008, Hosokawa:2011,Tanaka:2018, Fukushima:2020-7}, for simplicity. 
If the ionizing feedback reduces the mass supply from the envelope to the disc by the epoch we consider, it likely works against disc fragmentation. Our current results hence correspond to the maximal effects of the disc fragmentation for the cases of $\leq 10^{-5}$\:$\Zsun$.

%%%%%%%%%%%%%%%%%%%%%%%%%%%%%%%%%%%%%%%%%%%%
%%%%%%%%%%%%%%%%%%%%%%%%%%%%%%%%%%%%%%%%%%%%
%%% SECTION 3 %%%%
\section{Metallicity dependence of disc fragmentation}
\label{Sec:3}

%%%%%%%%%%%%%%%%%%%%%%%%%%%%%%%%%%%%%%%%%%%%
%%% SECTION 3-1 %%%%
\subsection{Disc fragmentation in 15 kyr since the disc formation}
\label{Sec:3-1}

%+--+--+--+--+--+--+--+--+--+--+--+--+--+--+--+--+--+--+--+--+%
\subsubsection{Fragmenting discs and self-gravitating clumps}
\label{Sec:3-1-1}
%+--+--+--+--+--+--+--+--+--+--+--+--+--+--+--+--+--+--+--+--+%

This section presents the simulation results for all the cases listed in \Tabref{Tab:2_LM} and clarifies the metallicity dependence of disc fragmentation. In particular, we study the masses and number of clumps forming via disc fragmentation at each metallicity.

%%%%% FIGURE 2 %%%%%
\begin{figure*}
 \begin{center}
 \begin{tabular}{c} 
  {\includegraphics[width=1.95\columnwidth]{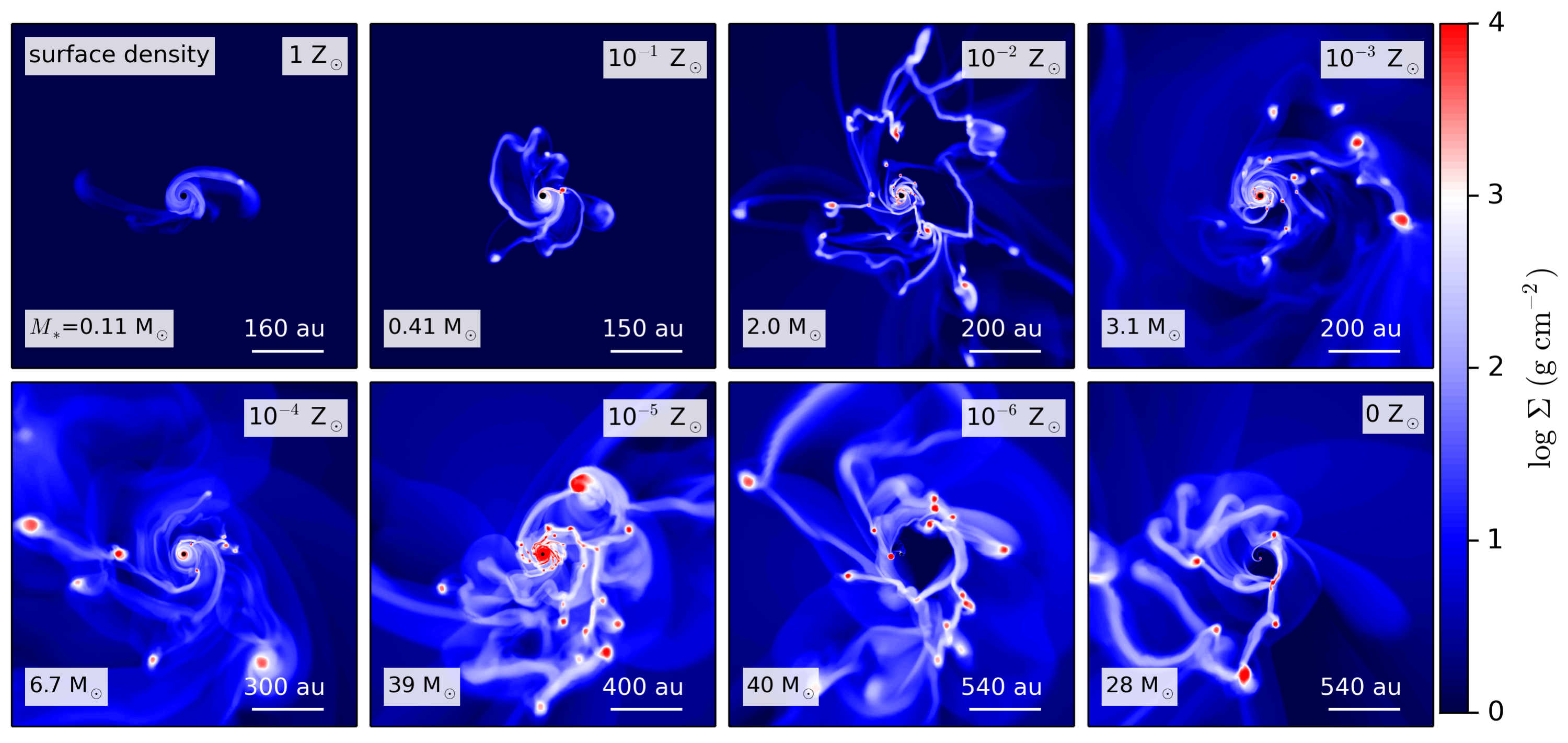}}
 \end{tabular}
 \caption{
 Spatial distributions of the surface density at the epoch of 15\:kyr after the disc formation. The panels represent the cases of the different metallicities listed in \Tabref{Tab:2_LM}. The mass of the central star is indicated in the lower-left corner of each panel. Note that the panel for the lower-metallicity case shows the broader area around the star. 
 }
 \label{Fig:Surface_1e-2}
 \end{center}
\end{figure*}
%

%---------------------------------------------------%

\Figref{Fig:Surface_1e-2} shows the spatial distribution of the surface density around the central star at $t = 15$\:kyr after the disc formation. We see the clumpy structure for all the cases, suggesting that the disc fragmentation generally occurs by the epoch of the presented snapshots. The number of the clumps depends on the metallicity. In the case of $1$\:$\Zsun$, for example, only one clump is embedded within the spiral arm $\sim 100{\rm\:au}$ away from the centre. 
On the other hand, at lower metallicities of $< 10^{-2}$\:$\Zsun$, there are a large number of clumps, which prevent the buildup of a well-defined disc structure. These results indicate that the disc fragmentation is more vigorous at lower metallicities.

%---------------------------------------------------------%

\Figref{Fig:Surface_1e-2} also shows that the spiral arms and clumps distribute over a broader area at lower metallicities. At $1$\:$\Zsun$, for example, the spatial extent of the spiral arms is limited to a small area of $\simeq 200$\:au, whereas it covers a larger size of $\sim1000$\:au at $0$\:$\Zsun$. These differences originate from higher prestellar cloud temperatures (or higher sound speeds) at lower metallicity (see \Tabref{Tab:2_LM}). The higher the sound speed, the broader range of the gas accretes onto the star-disc system by a given epoch, which provides a larger angular momentum. Therefore, the disc and spiral arms spread out faster with lower metallicities.

%-----------------------------------------------------------%

In general, high sound speed of the prestellar cloud $c_{\mathrm{s,0}}$ leads to a high infall rate from the envelope onto the central disc-star system (see also \Secref{Sec:3-2}), which is approximately described as 
\begin{align}
\dot{M}_{\rm infall} \simeq \frac{M_{\mathrm{J}}}{t_{\mathrm{ff}}} 
\simeq \frac{c_{\mathrm{s}}^{3}}{G},
\label{Eq:M_dot}
\end{align}
where $M_{\mathrm{J}} = 4/3 \pi\rho\left( l_{\mathrm{J}}/2 \right)^{3}$ is the Jeans mass, and $t_{\mathrm{ff}} = \sqrt{3\pi/\left(32G \rho\right)}$ is the free-fall time \citep{Shu:1977, Stahler:1986}.
Since the sound speed is higher at lower metallicity, the mean accretion rate onto the protostar $\dot{M}_{\ast}$ is also higher in our cases, because $\dot{M}_{\ast}$ is roughly proportional to $\dot{M}_{\rm infall}$.
The mass of the central star is consequently higher at lower metallicity at a given epoch. In fact, at the snapshots of $t=15{\rm\:kyr}$ presented in \Figref{Fig:Surface_1e-2}, the central stellar mass is $28$\:$\Msun$ for $0$\:$\Zsun$, while it is only $0.11$\:$\Msun$ for $1$\:$\Zsun$.

%%%%%% FIGURE 3 %%%%%
\begin{figure*}
 \begin{center}
 \begin{tabular}{c} 
  {\includegraphics[width=1.95\columnwidth]{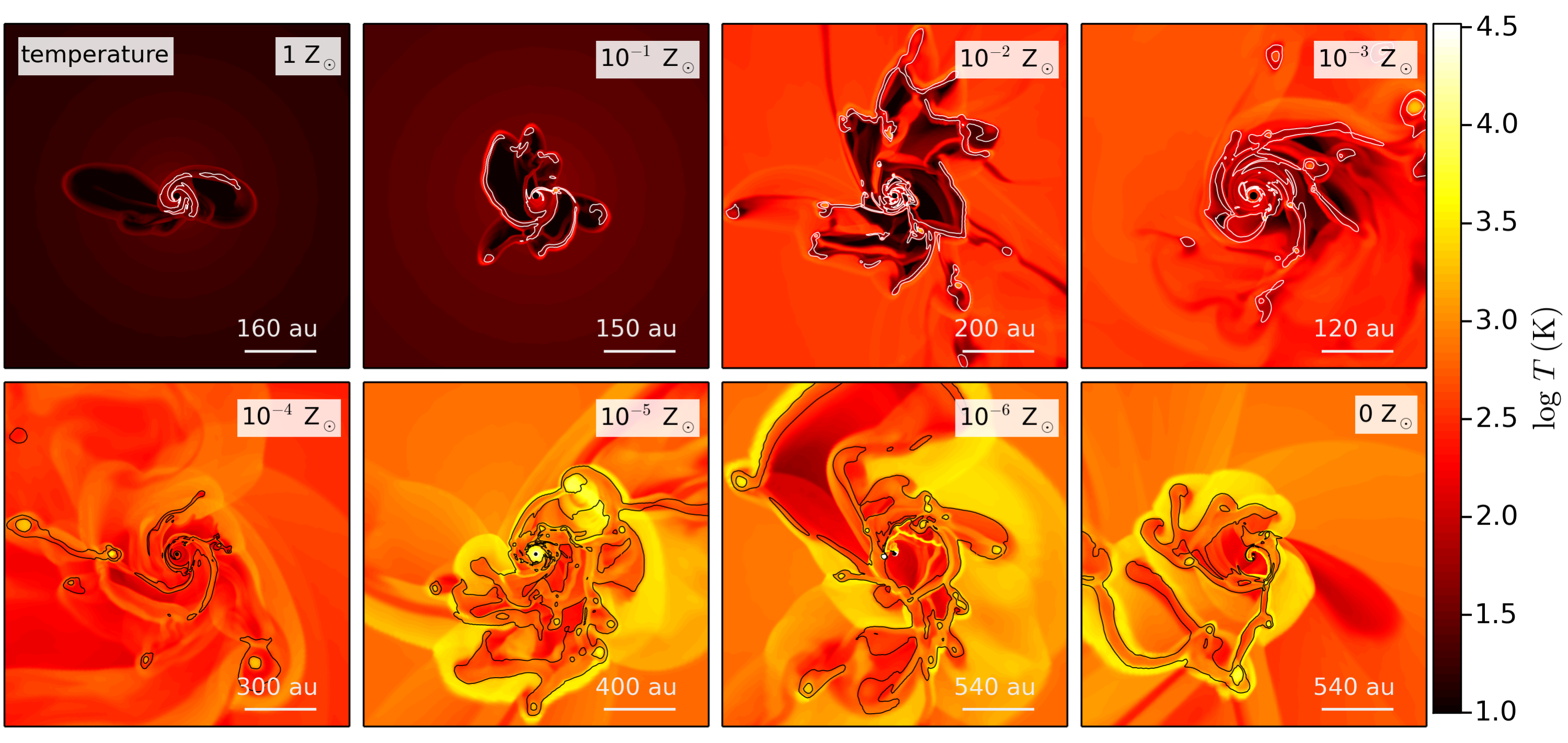}}
 \end{tabular}
 \caption{
Same as \Figref{Fig:Surface_1e-2} but for the spatial distributions of the gas temperature. The contour lines indicate iso-surface-density lines with $\log (\Sigma/{\rm g\:cm^{-2}})=1.5$ and $3.0$ (white in top panels, and black in bottom panels).}
 \label{Fig:Temperature_1e-2}
 \end{center}
\end{figure*}
%

%----------------------------------------------------------------------%

\Figref{Fig:Temperature_1e-2} shows the spatial distributions of the gas temperature at the same epoch as in \Figref{Fig:Surface_1e-2}. As with the surface density, we also find significant differences in the temperature structure among the models. The temperature tends to be higher for lower metallicity because the cooling becomes less efficient. At metallicities of $1$ and $0.1$\:$\Zsun$, the dust cooling is efficient everywhere within the regions presented in the panels. The energy transferred from the gas to dust via collisions
is dissipated by the dust thermal emission
(Equation~\ref{Eq:thermal_balance}).
In the envelope where the density is relatively low $<10^{6}$\:cm$^{-3}$, the temperature is $\sim10$\:K because of such efficient cooling (see also \Figref{Fig:nT_massgrid}). The temperature slightly rises to $\simeq30$--$100$\:K within $\sim100$\:au around the central star because the stellar irradiation is effective in addition to the compressional heating (see also \Secref{Sec:4-1} for more details). 
At $\le10^{-2}$\:$\Zsun$, the dust cooling becomes inefficient in the accretion envelope, where the temperature is mostly determined by the thermal balance between the energy loss by molecular-line emission and the compressional heating. The envelope temperature at $\le10^{-2}$\:$\Zsun$ is higher (>100\:K) than that at $\geq 0.1$\:$\Zsun$ as shown in \Figref{Fig:Temperature_1e-2}. 
In the cases of $\le10^{-2}$\:$\Zsun$, the dust cooling becomes effective only in dense disc materials near the star, where frequent gas-dust collisions occur. As a result, dense structures such as spiral arms with $\sim 10^{8}$--$10^{12}$\:cm$^{-3}$ are colder than in the accretion envelope. 
The threshold density above which the dust cooling is efficient increases with decreasing metallicity. 
The region where the dust is the primary coolant gradually shrinks and becomes limited to very vicinity of the central star. Eventually, at $\le 10^{-6}$\:$\Zsun$, the dust cooling is no longer effective, and the energy loss by H$_2$ molecular emission dominates everywhere in the computational domain. At these metallicities, the disc temperature is the highest ($>10^{3}$\:K) among the models.

%%%%% FIGURE 4 %%%%%
\begin{figure*}
 \begin{center}
 \begin{tabular}{c} 
  {\includegraphics[width=1.95\columnwidth]{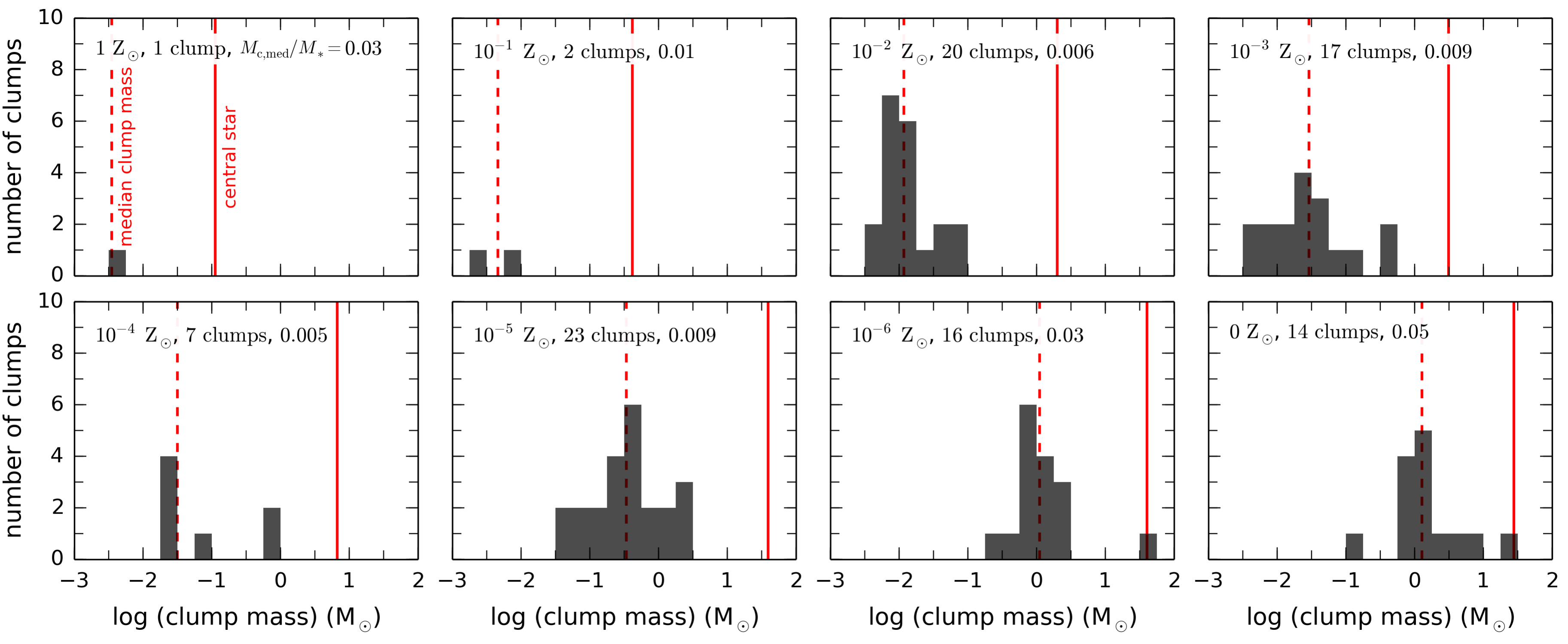}}
 \end{tabular}
 \caption{
 Mass distribution of clumps at 15\:kyr elapsed from the disc formation. The panels depict models with different metallicities indicated on the upper-left corners. In each panel, the red vertical lines represent the central stellar mass $M_{\ast}$ (solid) and the median mass of clumps $M_{\rm c,med}$ (dashed). The number of clumps and the ratio of the median mass to the central stellar mass are presented in the upper part of each panel. 
 }
 \label{Fig:MassDis_1e-2}
 \end{center}
\end{figure*}
%

%-----------------------------------------------------------------------------------%

\begin{figure*}
 \begin{center}
 \begin{tabular}{c} 
  {\includegraphics[width=1.95\columnwidth]{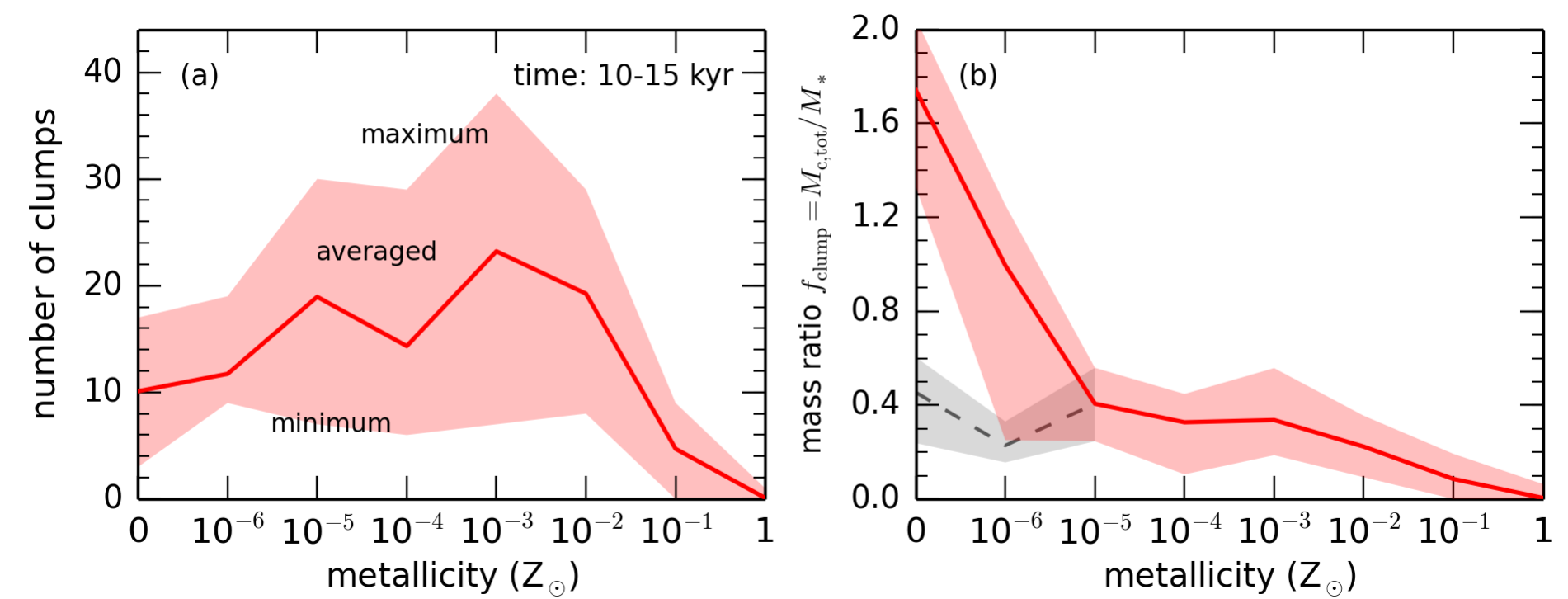}}
 \end{tabular}
 \caption{
 Metallicity dependence of (a) the number of clumps and (b) the mass ratio $f_{\mathrm{clump}}$, 
 i.e. the ratio of the sum of all clump masses $M_\mathrm{c,tot}$ to the central stellar mass $M_{\ast}$, at 10--15\:kyr from the disc formation. 
 In both panels, the solid red line represents the time-averaged value over 10--15\:kyr, 
 and the red shaded area depicts the range over which the quantities vary within the period.
 In panel (b), the black dashed line and grey shaded area represent the cases where the most massive clump is regarded as a part of the central star. That is, $f_{\mathrm{clump}}$ is redefined as $(M_\mathrm{c,tot} - M_\mathrm{c,massive})/(M_{\ast} + M_\mathrm{c,massive})$, where $M_\mathrm{c,massive}$ is the mass of the most massive clump (also see the text).
 }
 \label{Fig:Clump_1e-2}
 \end{center}
\end{figure*}

To examine physical properties of the clumps observed in the simulations quantitatively, we present their mass distribution at $t = 15$\:kyr in \Figref{Fig:MassDis_1e-2}. We describe the method to identify self-gravitating clumps separately in \Appref{App:frag}. 
\Figref{Fig:MassDis_1e-2} shows that the number of clumps is only 1 -- 2 for $1$\:$\Zsun$ and $10^{-1}$\:$\Zsun$,
while it is as large as $\sim10$--$20$ for $\leq 10^{-2}$\:$\Zsun$. We also show the metallicity dependence of the clump number over a longer duration 10--15\:kyr in \Figref{Fig:Clump_1e-2}a. We see the same overall trend as in \Figref{Fig:MassDis_1e-2}: the clump number is highest for the cases of $10^{-2}$--$10^{-5}$\:$\Zsun$, and the maximum mean value is $\sim 15$. Note that the clump number varies by almost an order of magnitude during the 5-kyr duration in these cases, as a result of clump merging, migration toward the star and local fragmentation. In \Figref{Fig:MassDis_1e-2}, for instance, the number of clumps at $10^{-4}$\:$\Zsun$ is significantly less than those for the adjacent $10^{-3}$ and $10^{-5}$\:$\Zsun$ cases.
However, \Figref{Fig:Clump_1e-2}a illustrates that this is just a transient phenomenon.
\Figref{Fig:Clump_1e-2}a also shows that the number of clumps decreases for $\gtrsim 10^{-2}$\:$\Zsun$ and $\lesssim 10^{-5}$\:$\Zsun$ toward both the low- and high-metallicity ends. Nevertheless, the decrease at $\lesssim 10^{-6}$\:$\Zsun$ is modest, and the time-averaged value is no less than $\sim10$. 

%------------------------------------------------------------------%

%%%%% FIGURE 5 %%%%%

%------------------------------------------------------------------------%

\Figref{Fig:MassDis_1e-2} suggests that not only the number of clumps but also their mass distribution depends on the metallicity: the typical clump mass increases with decreasing metallicity\footnote{
\Figref{Fig:MassDis_1e-2} also suggests that the minimum clump mass increases with decreasing metallicity, consistent with analytical argument on the gas giant planet formation via disc fragmentation by \cite{Johnson:2013}. 
}. 
At the metallicity of 1 and $0.1$\:$\Zsun$, the median clump mass is $M_{\rm c,med} < 10^{-2}$\:$\Msun$, while it exceeds $10^{-2}$\:$\Msun$ at $\simeq 10^{-2}$\:$\Zsun$ and rises to $\sim 1$\:$\Msun$ for $\lesssim 10^{-6}$\:$\Zsun$. While the overall trend is similar to the metallicity dependence of the central stellar mass, the median clump mass is generally smaller than the central stellar mass by more than one order of magnitude. The contrast is particularly large for $10^{-2}$--$10^{-4}$\:$\Zsun$, where the number of clumps is largest. The median clump mass is more than two orders of magnitude smaller than the stellar mass for these cases. From $1$\:$\Zsun$ toward lower metallicities, the median clump mass does not vary much from $\sim 10^{-2}$\:$\Msun$ until $\sim 10^{-4}$\:$\Zsun$, while the central stellar mass monotonically shifts to a larger value. We investigate the cause of this trend in \Secref{Sec:3-1-2}.

%+--+--+--+--+--+--+--+--+--+--+--+--+--+--+--+--+--+--+--+--+%
\subsubsection{Metallicity dependence of typical clump mass}
\label{Sec:3-1-2}
%+--+--+--+--+--+--+--+--+--+--+--+--+--+--+--+--+--+--+--+--+%

\begin{figure*}
 \begin{center}
 \begin{tabular}{c} 
  {\includegraphics[width=1.95\columnwidth]{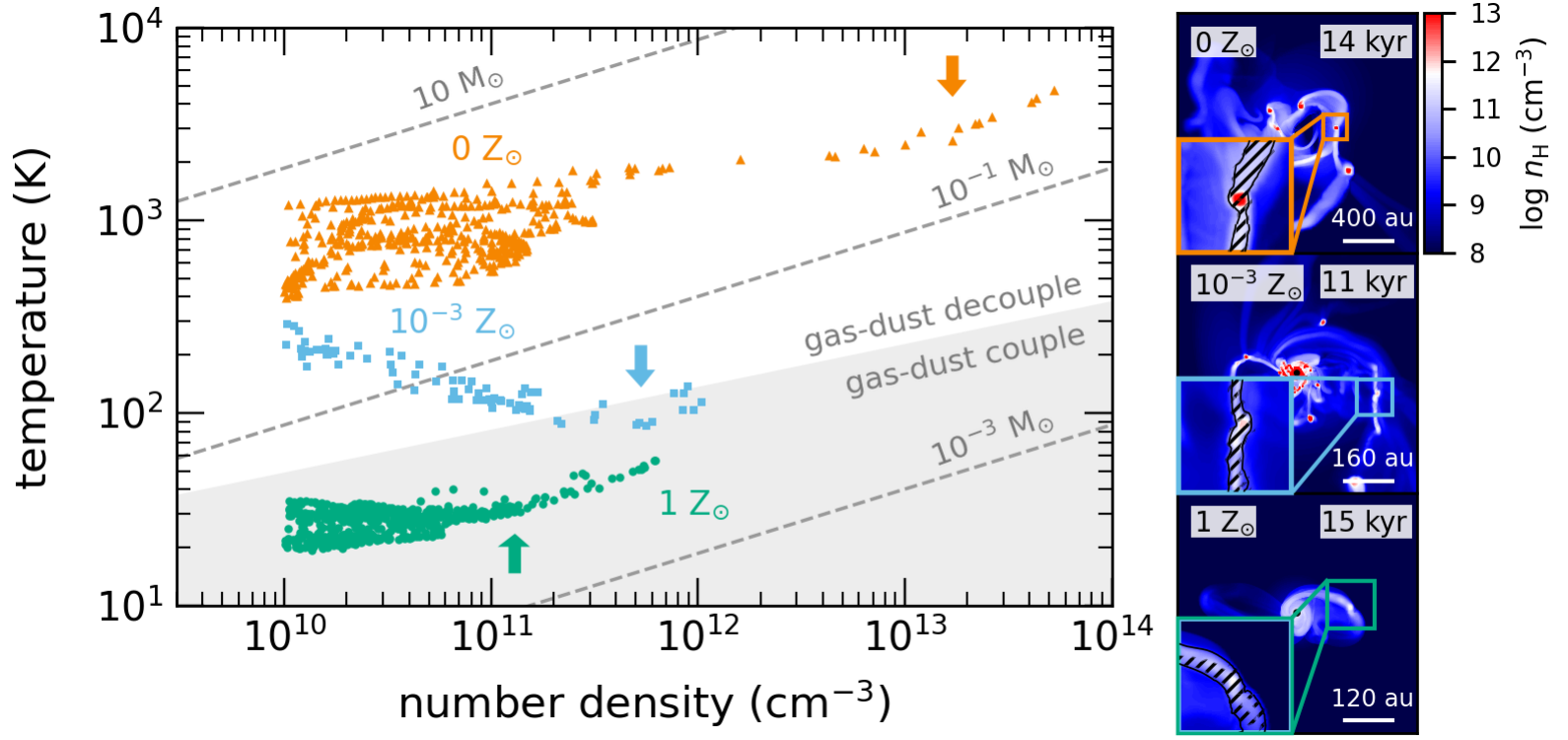}}
 \end{tabular}
 \caption{
 {\it Left panel}: 
 Metallicity dependence of the gas thermal state of the clumps forming in spiral arms and their surroundings. 
 The colours of the data points represent different metallicities, 
 $1$\:$\Zsun$ (green circles), $10^{-3}$\:$\Zsun$ (blue squares), and $0$\:$\Zsun$ (orange triangles). 
 The grey dashed line is the equal Jeans mass line, and the grey shaded area represents the state where the gas and dust temperature couple (also see text). 
 The arrows indicate the temperature turning points at which the Jeans mass is estimated. 
 {\it Right panels}:
 Snapshots of the number density distribution at the same epochs displayed in the left panel. 
 The metallicity and elapsed time at which the snapshot is taken are shown at the top of each panel. A magnified view around the clump is depicted in the lower-left corner of each panel.
 The area surrounding the clump with a number density of $> 10^{10}$\:cm$^{-3}$ corresponds to the gas whose thermal state is displayed in the left panel. 
 The black line marks the boundary of the region with the Toomre parameter $\mathcal{Q}_{\rm T} < 1$, and in the hatched regions the fragmentation criteria are satisfied; i.e, both $\mathcal{Q}_{\rm T} < 1$ and $\mathcal{G}<2\left[ 75\gamma(\gamma-1) \right]$, where $\mathcal{G}$ is the normalized cooling time (see Equations \ref{Eq:Toomre} and \ref{Eq:Gammie}).
 }
 \label{Fig:nT_1e-2}
 \end{center}
\end{figure*}

Recent theoretical studies show that the fragmentation of spiral arms is essential in driving the disc fragmentation \citep[e.g.][]{Takahashi:2016, Brucy:2021}. To understand the metallicity dependence of the typical clump mass, we consider the gas thermal evolution during the fragmentation of a spiral arm, and its metallicity dependence.
\Figref{Fig:nT_1e-2} shows gas thermal states of dense parts of spiral arms on the $n$-$T$ plane, where the fragmentation is about to take place for $1$, $10^{-3}$, and $0$\:$\Zsun$. As shown in the right panels, we pick up spiral arms that are somewhat isolated because numerous spiral arms form an intricate structure near the centre. Indeed, fragmentation is prone to occur in such outer disc parts, in agreement with studies on fragmentation of solar-metallicity discs \citep{Vorobyov:2015}. 

%------------------------------------------------------------------%

We confirm in those regions the conditions for the disc fragmentation
are satisfied; $\mathcal{Q}_{\rm T}<1$ \citep{Toomre:1964} and $\mathcal{G} < 2/\left[ 75\gamma(\gamma-1) \right]$ \citep{Gammie:2001,Rice:2005}, for which the Toomre parameter $\mathcal{Q}_{\rm T}$ and the normalized cooling time $\mathcal{G}$ are defined as
\begin{align}
\mathcal{Q}_{\rm T} = \frac{c_{\rm s}\Omega}{\pi G\Sigma} ,
\label{Eq:Toomre}
\end{align}
\begin{align}
\mathcal{G} = t_{\rm cool} \Omega ,
\label{Eq:Gammie}
\end{align}
where $\Omega$ is the Kepler angular velocity
and $t_{\rm cool}=e/Q_{\rm tot}$ is the cooling time. In the insets of the right panels we can see that both these conditions are satisfied in the fragmenting spiral arms.

%--------------------------------------------------------------------------%

In the left panel, the grey shaded area represents the state where the gas and dust temperatures well couple in a gravitationally collapsing cloud \citep{Tanaka:2014}: 
\begin{align}
T < 49~\left( \frac{n_{\mathrm{H}}}{10^{10}\:\mathrm{cm}^{-3}} \right)^{2/9}\:\mathrm{K}.
\end{align}
The collisional energy exchange between gas and dust becomes effective as the thermal state approaches this area. 
If the local density in the spiral arm becomes high enough to enter in the shaded area, the dust cooling affects the gas thermal state significantly.

%--------------------------------------------------------------------%

First, at $1$\:$\Zsun$ (green circles), all the data plots are within the shaded area, suggesting the efficient gas-dust energy exchange and then the dust cooling throughout the fragmentation process. 
The temperature is almost constant at $\simeq 20$--$30$\:K for the density of $\lesssim 10^{11}$\:cm$^{-3}$ and rises with the density for $\gtrsim 10^{11}$\:cm$^{-3}$. This temperature rise indicates that dust cooling becomes inefficient owing to enhanced optical depth, 
and the gas thermal evolution becomes adiabatic. The threshold density of $\sim 10^{11}$\:cm$^{-3}$ nearly corresponds to that for the so-called first hydrostatic cores (or Larson's first cores) in the present-day star formation \citep{Larson:1969,Omukai:2007,Tomida:2010,Saigo:2011}.

%--------------------------------------------------------------------%

Next, at $10^{-3}$\:$\Zsun$ (blue squares), the temperature drops from $\simeq 200$\:K to $60$\:K as the density increases from $10^{10}$ to $10^{12}$\:cm$^{-3}$. A striking feature of this case is that the data plots approach the shaded area with increasing density. The dust cooling is inefficient for $< 10^{10}$\:cm$^{-3}$, where the energy loss by molecular-line emission is the main cooling process. However, dust cooling becomes more and more effective as the density increases, so the temperature substantially decreases. After entering the shaded area for $\gtrsim 2\times10^{11}$\:cm$^{-3}$, the temperature turns to increase again. This occurs because, as in the case of $1$\:$\Zsun$, the dust radiative cooling becomes less effective owing to the photon trapping effect \citep{Tanaka:2014}. 
For this case, the gas distribution on the $n$-$T$ plane is the same as supposed for the so-called dust-induced fragmentation in the extremely metal-poor star formation \citep[e.g.][]{Omukai:2010}. Basically the same mechanism is responsible for the disc fragmentation.

%----------------------------------------------------------------------------%

Finally, at $0$\:$\Zsun$ (orange triangles), the data plots scatter in the upper part with $T \sim 10^{3}$\:K above the shaded area. 
With no dust grains in the primordial case, H$_2$ molecules are the dominant coolant for all the points. We see that the temperature increases above $\sim 10^{13}$\:cm$^{-3}$ to reach $T \simeq 5000$\:K. However, such behaviour is likely to be an artefact caused by our limited spatial resolution, rather than the photon trapping effect. We discuss this later in \Secref{Sec:4-2}.

%-----------------------------------------------------------------------------%

As shown above, \Figref{Fig:nT_1e-2} suggests that thermal evolution during the fragmentation of the arms varies with different metallicities. If the spiral-arm fragmentation is approximated as filament fragmentation, the Jeans mass at the local minimum on the $n$-$T$ plane should appear as a characteristic scale because the critical effective adiabatic index above that further contraction of filaments is prohibited is unity for such configuration. 
In \Figref{Fig:nT_1e-2}, the densities where the temperature turns to increase again, which are indicated by the arrows, are $\sim 10^{11}$\:cm$^{-3}$ for $1$\:$\Zsun$, $\sim 5 \times 10^{11}$\:cm$^{-3}$ for $10^{-3}$\:$\Zsun$, and $\sim 10^{13}$\:cm$^{-3}$ for $0$\:$\Zsun$.
The Jeans masses at these points are $\sim 10^{-2}$\:$\Msun$ for $1$ and $10^{-3}$\:$\Zsun$, and $\sim 1$\:$\Msun$ for $0$\:$\Zsun$. 
These values explain well the metallicity dependence of the median clump masses shown in \Figref{Fig:MassDis_1e-2}. 

%--------------------------------------------------------------------------%

%%%%% FIGURE 6 %%%%%

%+--+--+--+--+--+--+--+--+--+--+--+--+--+--+--+--+--+%
\subsubsection{Metallicity dependence of the clump number}
%+--+--+--+--+--+--+--+--+--+--+--+--+--+--+--+--+--+%

%%%%% FIGURE 7 %%%%%
\begin{figure}
 \begin{center}
 \begin{tabular}{c} 
  {\includegraphics[width=0.95\columnwidth]{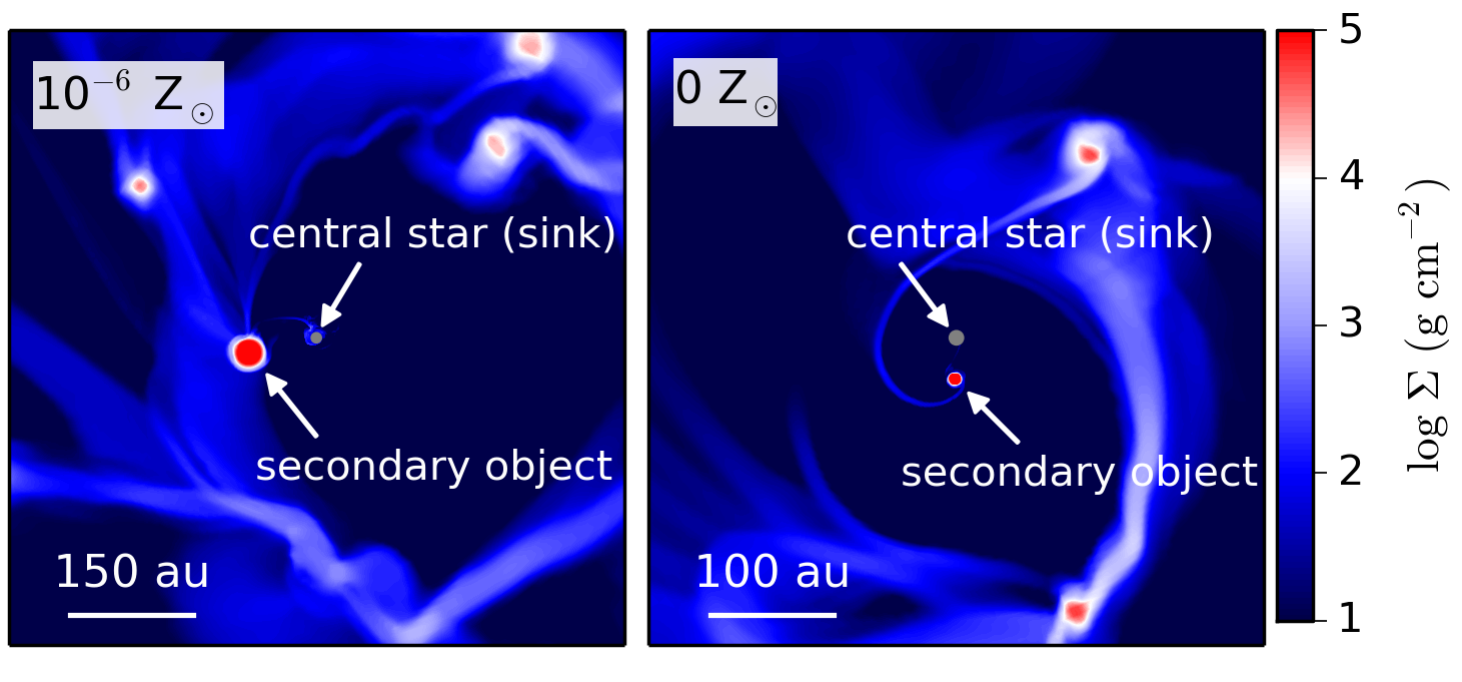}}
 \end{tabular}
 \caption{
 Enlarged views near the central star for the cases of $10^{-6}$\:$\Zsun$ (left) and $0$\:$\Zsun$ (right). The panels show the spatial distributions of the surface density over much smaller areas than in \Figref{Fig:Surface_1e-2}. The arrows mark the central star and most massive clump found in the snapshots. They are equally massive, forming long-lasting binary systems. 
 }
 \label{Fig:Surface_binary}
 \end{center}
\end{figure}

We interpret the metallicity dependence of the clump numbers (\Figref{Fig:Clump_1e-2}a) based on our consideration in \Secref{Sec:3-1-2}. For this purpose, we investigate the total mass of all the clumps $M_{\rm c,tot}$. \Figref{Fig:Clump_1e-2}b shows the metallicity dependence of the mass ratio defined as $f_{\rm clump} = M_{\rm c,tot}/M_{\ast}$.
From \Figref{Fig:Clump_1e-2}b, we see that $f_{\rm clump}$ is larger at lower metallicities. The mass ratio $f_{\rm clump}$ is smaller than 0.1 at $1$--$0.1$\:$\Zsun$ because only a few clumps form. At $10^{-2}$--$10^{-5}$\:$\Zsun$, where the fragmentation is intense, $f_{\rm clump}$ is almost constant at $\simeq 0.2$--$0.4$. At $10^{-6}$ and $0$\:$\Zsun$, $f_{\rm clump}$ is apparently higher ($>1.0$) than at higher metallicities. However, these are peculiar cases as shown in \Figref{Fig:Surface_binary}: a clump as massive as the central star orbits around the central star in the small separations of $< 100$\:au. 
The most massive clump, i.e., the secondary object, accretes the gas at a higher rate than the central star after its emergence (see also \Secref{Sec:3-2}) and stays in a stable orbit for a long period. Compared to the other clumps, the secondary object is unique and should be treated in the same way as the central star. We accordingly redefine $f_{\rm clump}$ as $(M_{\rm c,tot}-M_{\rm c,massive})/(M_{\ast}+M_{\rm c,massive})$ only for these cases, where $M_{\rm c,massive}$ is the mass of the secondary object. 
\Figref{Fig:Clump_1e-2}b shows that with the new definition $f_{\rm clump} \simeq 0.2$--$0.5$ at $10^{-6}$ and $0$\:$\Zsun$ (grey filled region), which are comparable to the values for $10^{-2}$--$10^{-5}$\:$\Zsun$. In other words, $f_{\rm clump}$ is nearly constant regardless of the metallicity for the cases of $\leq 10^{-2}$\:$\Zsun$, where the disc fragmentation forms more than ten clumps.

%----------------------------------------------------------------------------------%

Recall that the median clump mass $M_{\rm c,med}$ is the smallest against the central stellar mass $M_{\ast}$ for the cases of $10^{-2}$--$10^{-4}$\:$\Zsun$ (\Figref{Fig:MassDis_1e-2}). Since the mass ratio $f_{\rm clump}$ is nearly constant as shown in \Figref{Fig:Clump_1e-2}b, the median clump mass $M_{\rm c,med}$ is smallest relative to the clump total mass $M_{\rm c,tot}$ for these cases. Therefore, the number of clumps $\sim M_{\rm c,tot}/M_{\rm c,med}$ is largest at these metallicities, the trend shown in \Figref{Fig:Clump_1e-2}a. The above discussion suggests that the slight decline of the clump number at $\leq 10^{-5}$\:$\Zsun$ (\Figref{Fig:Clump_1e-2}a) is due to the increase of the typical clump mass. As mentioned above, however, this is likely an artificial trend caused by the limited spatial resolutions (see also \Secref{Sec:4-2}).

%%%%%%%%%%%%%%%%%%%%%%%%%%%%%%%%%%%%%%%%%%%%
%%% SECTION 3-2 %%%%
\subsection{Disc fragmentation in early period before 10 kyr}
\label{Sec:3-2}

\Secref{Sec:3-1} has described the disc fragmentation from 10 to 15\:kyr after the disc formation.
Next, we investigate the clump number and accretion rates onto the central star, including the earlier epoch than $t=10$--$15$\:kyr.

%+--+--+--+--+--+--+--+--+--+--+--+--+--+--+--+%
\subsubsection{Number of clumps increasing or decreasing with time}
%+--+--+--+--+--+--+--+--+--+--+--+--+--+--+--+%

%%%%% FIGURE 8 %%%%%
\begin{figure}
 \begin{center}
 \begin{tabular}{c} 
  {\includegraphics[width=0.95\columnwidth]{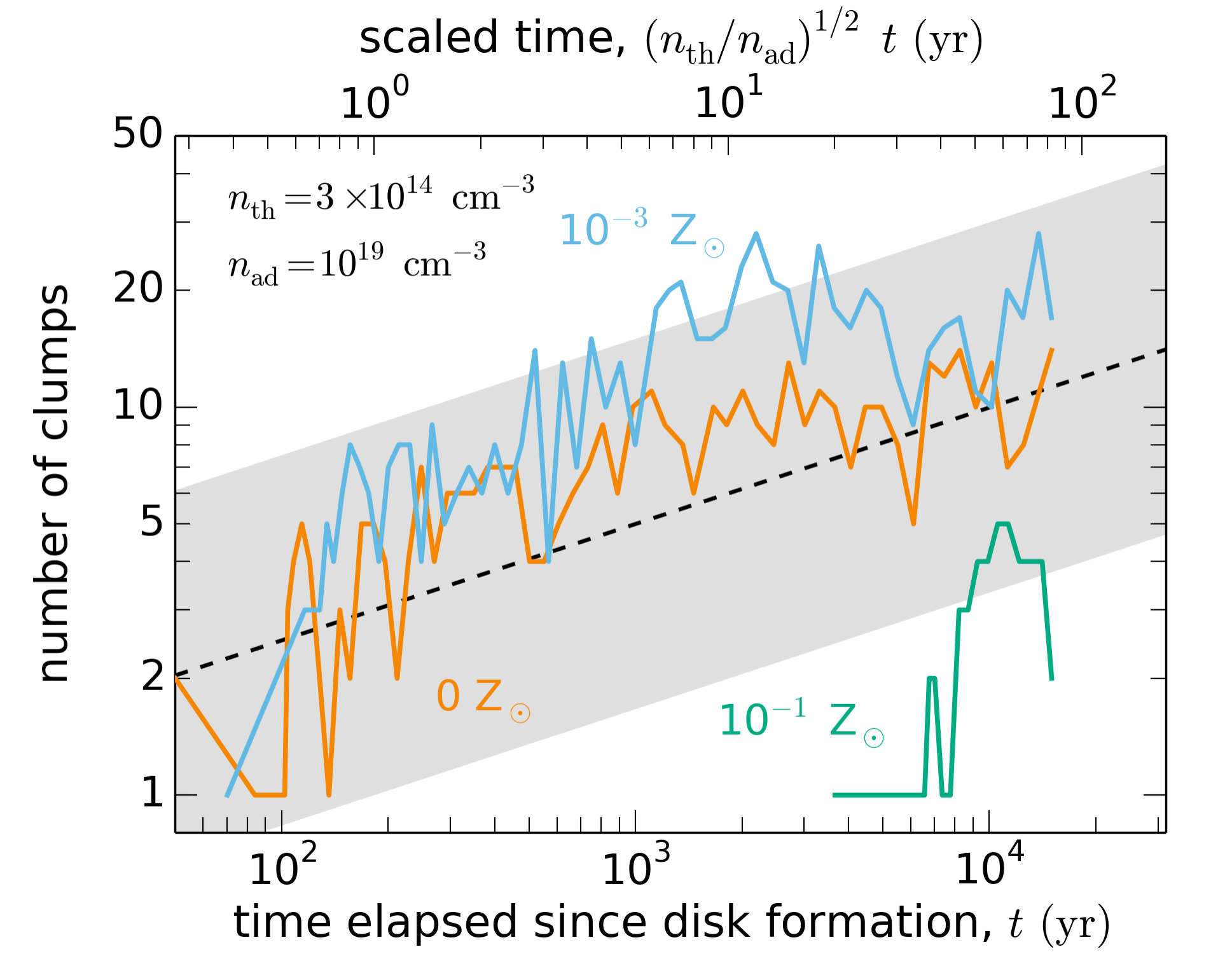}}
 \end{tabular}
 \caption{
 Metallicity-dependent evolution of the number of clumps for 15\:kyr after the epoch of disc formation. The line colours represent different metallicities, $10^{-1}$ (green), $10^{-3}$ (blue), and $0$\:$\Zsun$ (orange). The lower and upper horizontal axes are the elapsed time since the disc formation and the time normalized as $\sqrt{n_{\mathrm{th}}/n_{\mathrm{ad}}}t$, 
 where $n_{\mathrm{th}}=3\times10^{14}$\:cm$^{-3}$ and $n_{\mathrm{ad}}=10^{19}$\:cm$^{-3}$. 
 The black dashed line represents the scaling relation \Eqref{Eq:Susa2019} given by \citet{Susa:2019}. The grey shaded area denotes the range bounded by 3 and 1/3 times the black dashed line, where previous simulation results of the primordial star formation distribute \citep{Susa:2019}.
 }
 \label{Fig:Susa_1e-2}
 \end{center}
\end{figure}
%

%-------------------------------------------------------------------%

\Figref{Fig:Susa_1e-2} shows the time evolution of the number of clumps for the cases of $10^{-1}$, $10^{-3}$, and $0$\:$\Zsun$. Even in the early stage of $t<10$\:kyr, the metallicity dependence of the clumps number is almost the same as in \Figref{Fig:Clump_1e-2}a.
The clump number at $10^{-1}$\:$\Zsun$ is always the smallest among those three cases.
The first clump appears late at $\simeq 3$\:kyr, and the clump number is five even at the maximum. In contrast, at $10^{-3}$ and $0$\:$\Zsun$, the first clump appears shortly (<100\:yr) after the disc formation, and then the clump number continuously increases to $\sim 10$--$20$ with some oscillation in number. 

%-------------------------------------------------------------------%

As shown in \Figref{Fig:nT_1e-2}, at metallicities $1$ and $10^{-3}$\:$\Zsun$, our simulations resolve characteristic densities above which the gas becomes optically thick and adiabatic, i.e. $n_{\rm ad}\sim10^{11}$--$10^{12}$\:cm$^{-3}$. At $0$\:$\Zsun$, however, the gas becomes adiabatic at a very high density of $10^{19}$\:cm$^{-3}$ \citep[e.g.][]{Omukai:1998}, 
which is unresolved in most simulations of the disc fragmentation, including ours (see \Figref{Fig:nT_1e-2}). One must be careful when comparing the simulation results with different resolutions. 
Previous studies show that lower resolution leads to less disc fragmentation and fewer clumps with higher masses because the threshold density above which the gas becomes adiabatic is lower than $\sim 10^{19}$\:cm$^{-3}$ in a resolution-dependent way.
Nonetheless, \cite{Susa:2019} found an empirical relation between the clump number $N_{\rm c}$ and the time elapsed since the protostar formation $\Delta t$:
\begin{align}
N_{\mathrm{c}} \simeq 3\left[ \sqrt{\frac{n_{\mathrm{th}}}{n_{\mathrm{ad}}}} 
\left(\frac{\Delta t}{1\:\mathrm{yr}}\right) \right]^{0.3},
\label{Eq:Susa2019}
\end{align}
which integrates previous simulation results with different resolutions. The threshold $n_{\rm th}$ represents the density at the resolution limit. For our simulations, we estimate $n_{\mathrm{th}}=3\times10^{14}$\:cm$^{-3}$, with which the Jeans length assuming $T=10^{3}$\:K becomes equal to the sink radius, following  \cite{Susa:2019}. 
We compare our result of $0$\:$\Zsun$ to \Eqref{Eq:Susa2019}, using the time measured from the disc formation $t$ for $\Delta t$. \Figref{Fig:Susa_1e-2} shows that our result agrees with \Eqref{Eq:Susa2019} quite well, supporting the argument by \cite{Susa:2019}. We can see similar evolution also in our case of $10^{-3}$\:$\Zsun$, with always slightly high values. We further discuss our results, comparing them with previous studies in \Secref{Sec:4-4}.

%+--+--+--+--+--+--+--+--+--+--+--+--+--+--+--+--+--+--+--+%
\subsubsection{Mass accretion histories onto the central star}
%+--+--+--+--+--+--+--+--+--+--+--+--+--+--+--+--+--+--+--+%

%%%%% FIGURE 9 %%%%%
\begin{figure*}
 \begin{center}
 \begin{tabular}{c} 
  {\includegraphics[width=1.70\columnwidth]{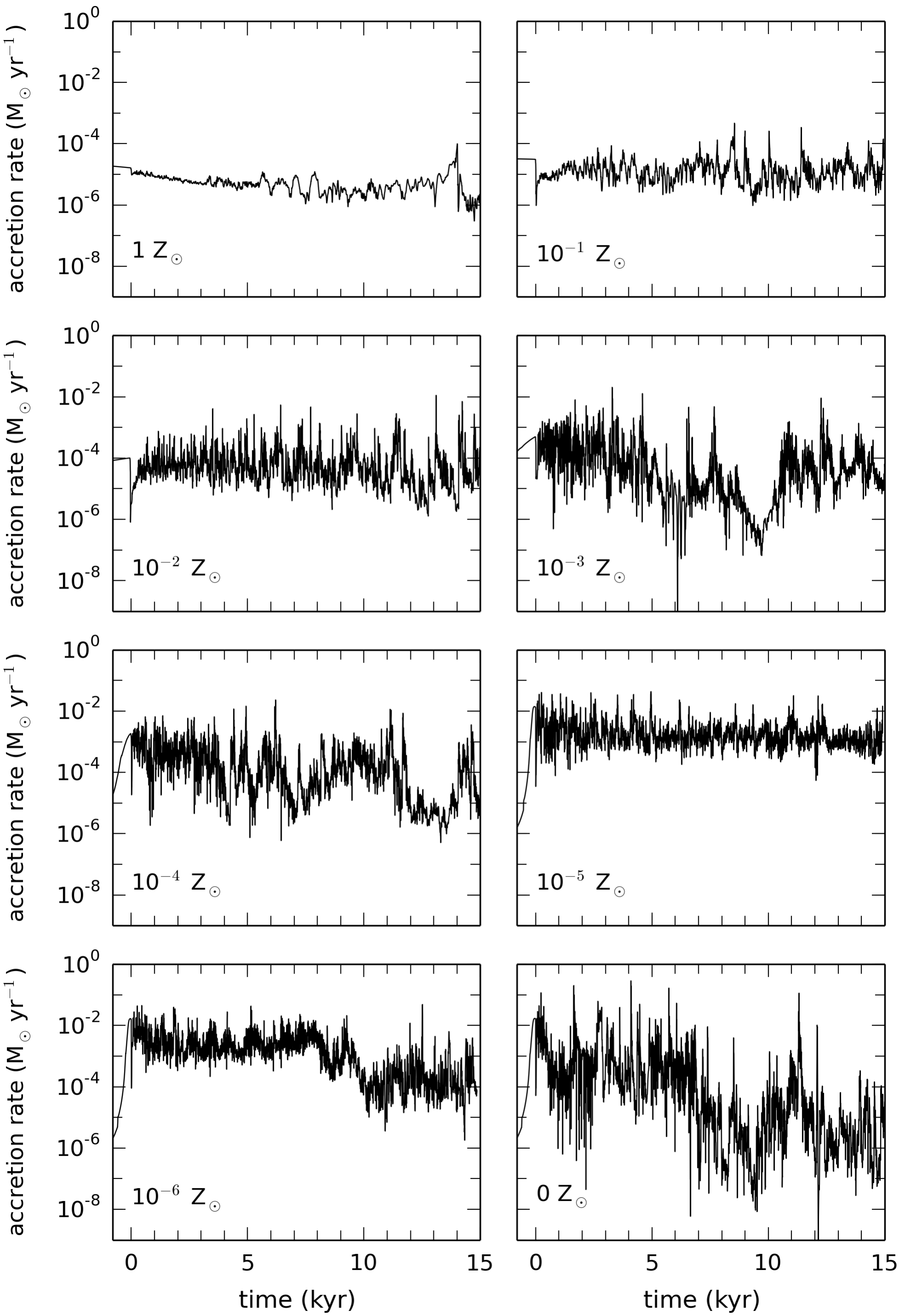}}
 \end{tabular}
 \caption{
 Time evolution of the accretion rate onto the central star from the epoch of disc formation ($t=0$) to $t=15$\:kyr. The panels represent the cases with different metallicities, as indicated in the lower-left corners of the panels. 
 }
 \label{Fig:accretion_rate}
 \end{center}
\end{figure*}
%

%------------------------------------------------------------------------%

\Figref{Fig:accretion_rate} shows the time evolution of the mass accretion rates onto the central star until $t = 15$\:kyr for the different metallicities. 
As mentioned in \Secref{Sec:3-1-1}, the time-averaged accretion rate is higher for lower metallicities: whereas the mean rate is $\sim 10^{-5}$\:$\Msun$\:yr$^{-1}$ at $1$\:$\Zsun$, it is $\sim 10^{-3}$\:$\Msun$\:yr$^{-1}$ at $10^{-5}$\:$\Zsun$.

%------------------------------------------------------------%

In \Figref{Fig:accretion_rate}, there are spiky features over a short period of $\ll1$\:kyr in all the cases. These violent fluctuations reflect temporary changes in the accretion rate due to the infall of clumps onto the central star. The frequency of these spiky fluctuations correlates with the number of clumps. For instance, the frequency increases as the metallicity decreases from $1$\:$\Zsun$ while the number of clumps increases. The typical amplitude of the fluctuations also increases with decreasing metallicity: whereas there are only mild fluctuations for $1$--$0.1$\:$\Zsun$, the accretion rate more vigorously varies for $\lesssim 10^{-2}$\:$\Zsun$.

%---------------------------------------------------------------%

The presence of clumps also causes long-term variability of the accretion rate over more than kyr, which is particularly noticeable for the cases of $10^{-6}$ and $0$\:$\Zsun$.
In these cases, the accretion rate onto the central star substantially drops at $t \simeq 10$\:kyr for $10^{-6}$\:$\Zsun$ ($t \simeq 7$\:kyr for $0$\:$\Zsun$), and it stays at the low values after that. The formation of the massive secondary object illustrated in \Figref{Fig:Surface_binary} causes such evolution. The secondary object has the typical fragment mass of $\sim 1$\:$\Msun$ at its birth (\Secref{Sec:3-1-2}), and it grows comparably massive to the central star ($\sim 25$--$30$\:$\Msun$) by accreting the gas from the envelope exclusively. As a result, the accretion rate onto the central star stays low during that period. 

%--------------------------------------------------------------%

In \Figref{Fig:accretion_rate}, the accretion rate significantly varies by about three orders of magnitude for a few kyr around $t = 10$ and 13\:kyr at $10^{-3}$ and $10^{-4}$\:$\Zsun$, respectively.
In these cases, the infalling gas temporarily accumulates at a region in the envelope with a density of $\sim 10^{6}$--$10^{7}$\:cm$^{-3}$.
\Figref{Fig:nT_massgrid} shows that efficient H$_2$-formation heating creates pressure excess around that part, which decelerates the infall motion \citep{Omukai:2010}.
We can also confirm this effect from \Figref{Fig:Surface_Init}b, where the surface density sharply changes at a radius of $\simeq 2000$--$4000$\:au for these metallicity cases. The accumulated gas falls again toward the centre in a free-fall manner after the density is elevated to $\sim 10^{8}$\:cm$^{-3}$. The corresponding free-fall timescale is $\simeq 4\:(n_{\rm H}/10^{8}\:{\rm cm}^{-3})^{-1/2}$\:kyr, which is comparable to the timescale of the accretion-rate variation seen in \Figref{Fig:accretion_rate}. The mass supply rate from the envelope to the disc fluctuates on the free-fall time, and the accretion rate onto the central star is also variable.

%%%%%%%%%%%%%%%%%%%%%%%%%%%%%%%%%%%%%%%%%%%%
%%%%%%%%%%%%%%%%%%%%%%%%%%%%%%%%%%%%%%%%%%%%
%%% SECTION 4 %%%%
\section{Discussion}
\label{Sec:4}

%%%%%%%%%%%%%%%%%%%%%%%%%%%%%%%%%%%%%%%%%%%%
%%% Section 4.1 %%%%
\subsection{Effects of protostellar irradiation}
\label{Sec:4-1}

%----------------------------------------------------------------%

Our simulations incorporate the effect of stellar irradiation heating, which has often been neglected in previous studies. In general, irradiation heating is more effective for higher-metallicity discs where the temperature is lower owing to the more efficient dust cooling (Equation \ref{Eq:thermal_balance} and \Figref{Fig:Temperature_1e-2}), 
and it works against the disc fragmentation. 
As reference models for comparison, we here perform trial simulations ignoring the irradiation heating for the cases of $1$, $10^{-1}$, and $10^{-2}$\:$\Zsun$. 
We consider the spatially-uniform background radiation field with 10\:K also in these cases. 
The initial density and angular velocity distributions are the same as in the standard runs described in \Secref{Sec:2-2}.

%%%%% FIGURE 10 %%%%%
\begin{figure*}
 \begin{center}
 \begin{tabular}{c} 
  {\includegraphics[width=1.97\columnwidth]{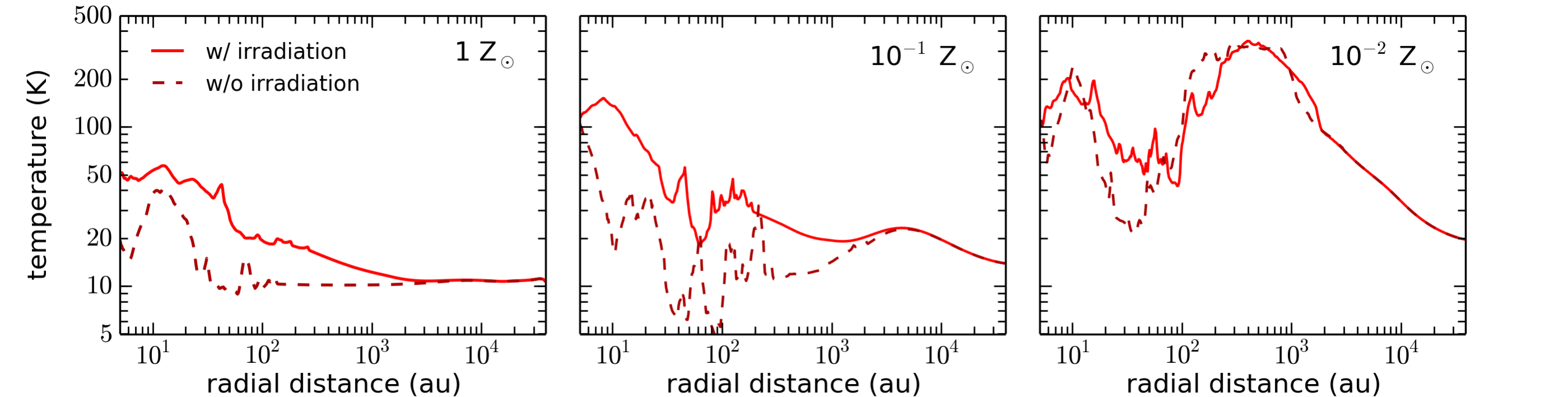}}
 \end{tabular}
 \caption{
Impact of stellar irradiation heating on the thermal structure of the disc and envelope with different metallicities.
The panels show the radial distributions of azimuthally-averaged temperature with different metallicities, $1$ (left), $10^{-1}$ (middle), and $10^{-2}$\:$\Zsun$ (right), at the same epoch as in \Figref{Fig:Temperature_noirr}.
The solid and dashed lines represent the cases with and without the irradiation heating in each panel, respectively. 
}
 \label{Fig:Temperature_radial}
 \end{center}
\end{figure*}
%

%-----------------------------------------------------------------------------%

\Figref{Fig:Temperature_radial} shows the azimuthally-averaged temperature distributions with and without the irradiation heating. 
For the cases of $1$\:$\Zsun$ and $10^{-1}$\:$\Zsun$, the irradiation heating is effective within a radius of $\simeq 3000$ au from the centre, enhancing the temperature by a factor of a few. At $10^{-2}$\:$\Zsun$, the irradiation heating is less prominent than in the above cases and slightly increases the temperature within $\sim 100$\:au from the centre.

%%%%% FIGURE 11 %%%%%
\begin{figure*}
 \begin{center}
 \begin{tabular}{c} 
  {\includegraphics[width=1.55\columnwidth]{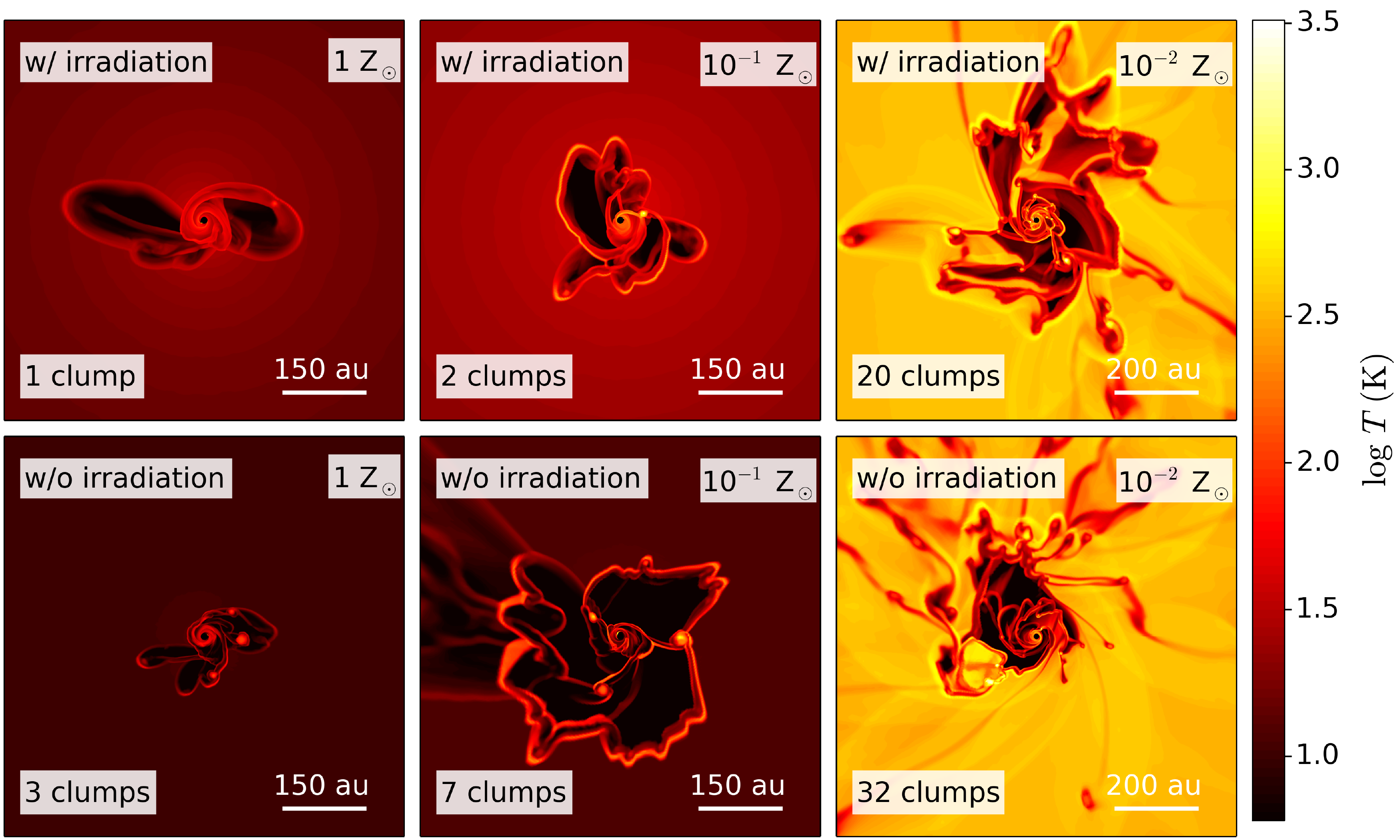}}
 \end{tabular}
 \caption{
Comparison of the temperature distributions around the central star with and without the central stellar irradiation (upper and lower panels). Each column represent a different metallicity, $1$, $10^{-1}$, and $10^{-2}$\:$\Zsun$, from left to right. All the panels show the snapshots taken at the epoch of 15\:kyr after the disc formation. The number of clumps is shown in the lower-left corner of each panel.
 }
 \label{Fig:Temperature_noirr}
 \end{center}
\end{figure*}
%

%----------------------------------------------------------------------------%

\Figref{Fig:Temperature_noirr} illustrates the two-dimensional spatial distributions of the temperature in the same cases. We clearly see that the stellar irradiation increases the temperature in the envelope and disc in the cases of $1$\:$\Zsun$ and $10^{-1}$\:$\Zsun$. 
By contrast, at $10^{-2}$\:$\Zsun$, its effect is less evident, affecting only the inner part of the disc.
\Figref{Fig:Temperature_noirr} also reveals that the irradiation heating stabilizes the disc and reduces the number of clumps, as reported by \cite{Bate:2012} and \cite{Stamatellos:2012} for solar-metallicity discs. 
This effect is particularly remarkable for the cases with $1$ and $10^{-1}$\:$\Zsun$. Without stellar irradiation, the number of clumps increases by a factor of about three. On the other hand, at $10^{-2}$\:$\Zsun$, where the temperature change is small, the difference in the number of clumps is also modest, with only 1.6 times.

%----------------------------------------------------------------------------%

While the stellar irradiation affects the disc fragmentation, it only plays a secondary role in shaping the overall metallicity dependence shown in \Figref{Fig:Clump_1e-2}a. 
Even without the stellar irradiation, the number of clumps decreases with increasing metallicity from $10^{-2}$\:$\Zsun$ (\Figref{Fig:Temperature_noirr}), which is the same trend as in \Figref{Fig:Clump_1e-2}a. 
The irradiation heating effectively stabilizes the disc at higher metallicities, and thus causes the clump number to depend more strongly on the metallicity.

%%%%%%%%%%%%%%%%%%%%%%%%%%%%%%%%%%%%%%%%%%%%
%%% Section 4.2 %%%%
\subsection{Limitation by spatial resolution and in search of ``true'' clump numbers}
\label{Sec:4-2}

%-----------------------------------------------------%

As described in \Secref{Sec:2-1}, we artificially suppress the cooling rate in our simulations (Equation~\ref{Eq:Climit}). While this method can prevent spurious fragmentation in under-resolved regions, there is a pitfall: it potentially introduces resolution dependencies in the results. 
We must be careful.
Calculations with sufficiently high resolution should mitigate such dependencies, but achieving it for all the cases is computationally infeasible. 
Hence, we here perform additional simulations only for representative cases $1$ and $10^{-3}$\:$\Zsun$ to examine the effect of the limited spatial resolution on disc fragmentation. 
While we have used 512$^2$ grid cells for the standard runs described in \Secref{Sec:3}, we here also examine the runs with 256$^2$, 768$^2$, and 1024$^2$ cells. We follow the evolution for 15\:kyr after the disc formation for all the additional runs.

%%%%% FIGURE 12 %%%%%
\begin{figure}
 \begin{center}
 \begin{tabular}{c} 
  {\includegraphics[width=0.95\columnwidth]{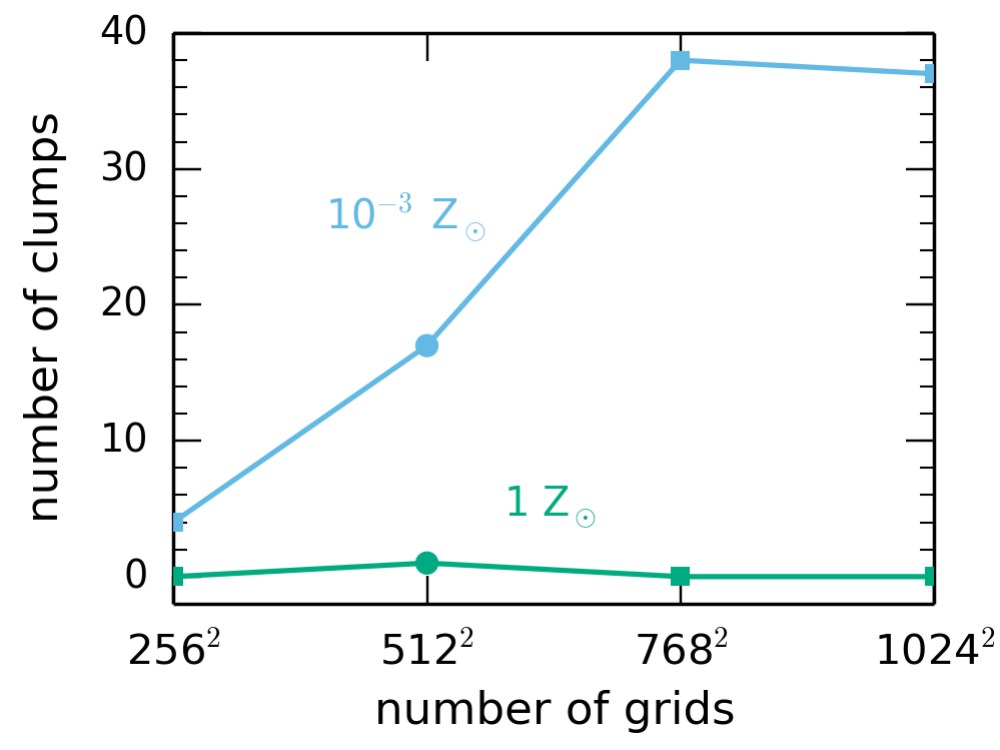}}
 \end{tabular}
 \caption{
 Effect of varying spatial resolution on the number of clumps at $t = 15$\:kyr after the disc formation. The circle and square symbols represent the standard cases using 512$^2$ grid cells and trial cases adopting different resolutions with 256$^2$, 768$^2$, and 1024$^2$ cells. The colours indicate different metallicities, $1$ (green) and $10^{-3}$\:$\Zsun$ (blue). 
 }
 \label{Fig:Clump_grid}
 \end{center}
\end{figure}

%-----------------------------------------------------------------------%

\Figref{Fig:Clump_grid} shows variations of the clump number with different spatial resolutions. We only see little resolution dependence for the cases of $1$\:$\Zsun$: the disc fragmentation hardly occurs even if we vary the resolution. However, the cases of $10^{-3}$\:$\Zsun$ show a large variation with the different resolutions. 
At this metallicity, the number of clumps monotonically rises from $\sim 5$ to $\sim 40$ as the grid number increases from 256$^2$ to 768$^2$. The results with 768$^2$ and 1024$^2$ cells are almost the same. 
\Figref{Fig:Clump_grid} thus suggests that our standard run with 512$^2$ cells underestimates the clump number at $10^{-3}$\:$\Zsun$.
Recall that the number of clumps tends to increase with decreasing metallicity from $1$ to $10^{-3}$\:$\Zsun$ (\Figref{Fig:Clump_1e-2}a). 
Considering the resolution dependence, we expect that the actual metallicity dependence is steeper than that shown in \Figref{Fig:Clump_1e-2}a.

%--------------------------------------------------------------------------%

Let us consider the reason for the large resolution dependence in the case of $10^{-3}$\:$\Zsun$ in \Figref{Fig:Clump_grid}. 
As discussed in \Secref{Sec:3-1-2}, the gas thermal evolution in a fragmenting spiral arm determines the typical clump mass at each metallicity (\Figref{Fig:nT_1e-2}).
The characteristic scale set by the Jeans length at the onset of the adiabatic evolution is smaller at a lower metallicity. For example, it is $\sim10$\:au scale of the first hydrostatic core at 1\:$\Zsun$ and $\sim10^{-2}$\:au scale of the protostellar core at 0\:$\Zsun$.
Spatially resolving it becomes more difficult with decreasing metallicity.
Note that, although the 512$^2$ resolution is generally insufficient at $10^{-3}$\:$\Zsun$, the fragmenting spiral arm in Figure \ref{Fig:nT_1e-2} was sufficiently resolved (e.g. $4x_{\rm grid}<l_{\rm J}$), and thus the interpretation given in Section \ref{Sec:3-1-1} should be correct. 

%---------------------------------------------------------------------------%

In the most extreme case of primordial star formation ($0$\:$\Zsun$), the characteristic scale is that of a protostar, which is far below our resolution limit.
Thus, we expect stronger resolution dependence in this case than for $10^{-3}$\:$\Zsun$ shown in \Figref{Fig:Clump_grid}.
Nonetheless, we can estimate the actual number of clumps $N_{\rm c}$ using \Eqref{Eq:Susa2019}: substituting $n_{\rm th}=n_{\rm ad}=10^{19}$\:cm$^{-3}$ and $\Delta t=15000$\:yr, \Eqref{Eq:Susa2019} reads $N_{\rm c} \simeq 50$. 
This is five times larger than that for
the primordial case ($\simeq 10$, see \Figref{Fig:Clump_1e-2}a) and slightly greater than that for $10^{-3}$\:$\Zsun$ with the resolution of 768$^2$ cells (\Figref{Fig:Clump_grid}). 
Therefore, the moderately decreasing trend of $N_{\rm c}$ from $10^{-3}$\:$\Zsun$ to $0$\:$\Zsun$ (\Figref{Fig:Clump_1e-2}a) is likely attributable to the limited spatial resolution. 
Some flat or slightly increasing trend at $N_{\rm c} \simeq 40$--$50$ may replace it in reality.

%----------------------------------------------------------------------------%

Even with the limited resolution, we can extract the actual metallicity dependence of the disc fragmentation. 
In \Appref{App:parameter}, we present additional simulations starting from different initial conditions; rapidly rotating cloud cores (\Appref{App:Rapid}) and massive cores (\Appref{App:Massive}).
Modifying the initial condition results in the change in the radii where fragmentation occurs, which also changes the effects of the spatial resolution as the outer region has the lower resolution in our log-polar coordinate system.
Therefore, distinguishing the physical effects of the initial conditions and the numerical limitation of spatial resolution is sometimes difficult.
Nonetheless, we can discuss the metallicity dependence of the disc fragmentation for a given initial core model.
We show that similar metallicity dependence exists even for the different initial models in Figures~\ref{Fig:Clump_5e-2}a and \ref{Fig:Clump_massive}a (cf. \Figref{Fig:Clump_1e-2}a).

%%%%%%%%%%%%%%%%%%%%%%%%%%%%%%%%%%%%%%%%%%%%
%%% Section 4.3 %%%%
\subsection{Long-term evolution for 150 kyr}
\label{Sec:4-3}

In \Secref{Sec:3}, we have analyzed the metallicity dependence of disc fragmentation, particularly the number of clumps and their mass distribution until 15\:kyr. Here, we extend the calculation up to 150\:kyr in the same cases of $1$ and $10^{-3}$\:$\Zsun$ and discuss the metallicity dependence in later evolution. 
At 150\:kyr, more than 65 \% of material in the parent cloud still remains in the envelope, which continues to supply the mass to the disc.

%%%%% FIGURE 13 %%%%%
\begin{figure}
 \begin{center}
 \begin{tabular}{c} 
  {\includegraphics[width=0.95\columnwidth]{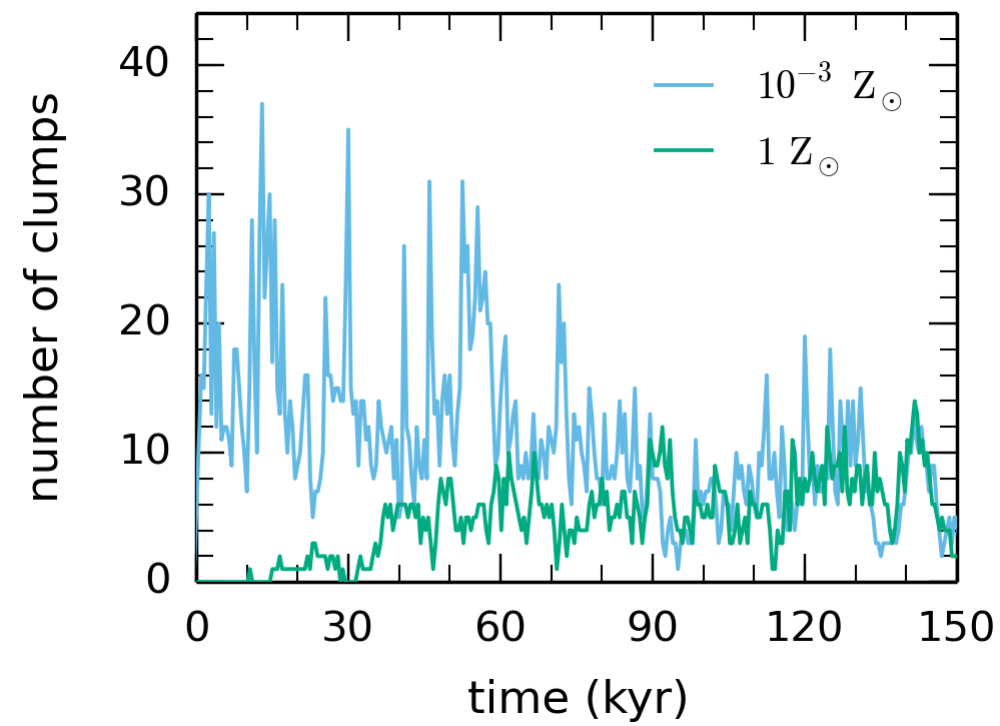}}
 \end{tabular}
 \caption{
 Long-term evolution of the number of clumps for 150\:kyr after the disc formation. 
 The green and blue lines show the cases of $1$\:$\Zsun$ and $10^{-3}$\:$\Zsun$, respectively. 
 }
 \label{Fig:Clump_long}
 \end{center}
\end{figure}
%

%----------------------------------------------------------------------------%

\Figref{Fig:Clump_long} shows the time evolution of the number of clumps in those cases for 150\:kyr. 
At 15\:kyr, the clump number is $\sim 10$ times larger with $10^{-3}$\:$\Zsun$ than with $1$\:$\Zsun$ (\Secref{Sec:3}), but their difference becomes smaller and eventually disappears after $\simeq 90$\:kyr. 
This apparently seems to indicate that the metallicity dependence of disc fragmentation described in \Secref{Sec:3} is only present in the early evolutionary stage. 
However, one has to be careful about interpreting the results. 

%---------------------------------------------------------------------------%

First, at $1$\:$\Zsun$, the first clump appears at $\sim10$\:kyr, and the number of clumps subsequently increases with some fluctuations. We can interpret this trend as follows. 
The disc radius becomes larger with time as the specific angular momentum of the infalling gas increases in our setting. As a result, the effect of the stellar irradiation heating becomes less effective at the outer edge of the disc, 
which is prone to fragmentation at $\geq 15$\:kyr (\Secref{Sec:4-1}).
In addition, the normalized cooling time $\mathcal{G}$ at the outer edge of the disc tends to be small because of low angular velocity. 
This promotes  fragmentation  and then the number of clumps increases \citep{Vorobyov:2020-6}.
\cite{Vorobyov:2010} and \cite{Vorobyov:2020-6} followed long-term disc evolution at $1$\:$\Zsun$ with very similar code and setting as ours, and reported active disc fragmentation consistent with our result.

%---------------------------------------------------------------------------%

Second, at $10^{-3}$\:$\Zsun$, dozens of clumps are formed immediately after the disc formation,
and the clump number gradually decreases with large fluctuations.
This clump-number decline can be interpolated as follows:
the density of the disc's outer edge decreases with time as the disc size extends,
and the dust cooling becomes inefficient with $\lesssim10^{10}$\:cm$^{-3}$, terminating the fragmentation.
However, we must see this result with particular attention to the numerical artefact. 
As discussed in \Secref{Sec:4-2}, the clump number depends on the spatial resolution at $10^{-3}$\:$\Zsun$. 
In the long-term evolution considered here, the disc size becomes larger, and the spatial resolution becomes effectively coarse with time in our log-polar coordinate system.
As shown in \Figref{Fig:Clump_grid}, the lack of resolution leads to underestimating the clump number,
which indicates that the decrease of the clump number at $10^{-3}$\:$\Zsun$ shown in \Figref{Fig:Clump_long} is partly due to the insufficient resolution. 
To confirm how long the metallicity dependence of the disc fragmentation observed in \Secref{Sec:3} persists during the accretion phase, one must systematically perform long-term simulations with sufficiently high resolutions. 
Such calculations are computationally demanding and outside the scope of this study. 
It belongs to research in the future world. 

%%%%%%%%%%%%%%%%%%%%%%%%%%%%%%%%%%%%%%%%%%%%
%%% Section 4.4 %%%%
\subsection{Paying tribute to our predecessors' achievements}
\label{Sec:4-4}

%---------------------------------------------------------------------------%

We here compare our simulation results with previous studies. 
By analyzing the gravitational instability of one-dimensional steady-state discs, 
\cite{Tanaka:2014} predicted that vigorous fragmentation occurs at $10^{-3}$--$10^{-5}$\:$\Zsun$.
This is indeed observed in our simulation, and a large number of clumps are produced in this metallicity range, qualitatively consistent with the claim by \cite{Tanaka:2014}. 

%---------------------------------------------------------------------------%

Intense disc fragmentation at low metallicities is also reported by some other authors who followed shorter-term evolution than ours \citep[$\lesssim1000$\:yr by ][]{Machida:2015, Shima:2021, Chiaki:2022}. 
In their calculation, the number of clumps  increases more rapidly than in ours, reaching $\sim 10$ no later than $\sim 100$\:yr after the birth of the protostar at $\leq 10^{-5}$\:$\Zsun$.
This can be attributed to the fact that their simulations resolve smaller scales ($\lesssim 10$\:au) with higher resolution. Our simulations may fail to follow such small-scale fragmentation because of the limited spatial resolution. 
In our long-term simulations, however, the disc fragmentation occurs in a later phase ($> 100$\:yr) producing $\sim 10$ clumps. 
Despite the difference in fragmentation epochs, both results indicate a similar tendency of frequent disc fragmentation at low metallicities. 

%---------------------------------------------------------------------------%

Some simulations of low-metallicity disc fragmentation followed longer-term evolution ($\sim10^{5}$\:yr) than ours \citep{Vorobyov:2020-9, Chon:2021-12}. 
They differ from our simulation in terms of the initial conditions, particularly the mass of initial cloud cores. 
Here, we vary the core mass as a function of the metallicity, $\simeq 3$--$1400$\:$\Msun$ for $1$--$0$\:$\Zsun$, the typical mass expected at each metallicity. 
On the other hand, in their studies, the initial mass was fixed in most cases; $\sim 1$\:$\Msun$ for $1$--$10^{-2}$\:$\Zsun$ in \cite{Vorobyov:2020-9}, and $\sim 1000$\:$\Msun$ for $10^{-1}$--$10^{-6}$\:$\Zsun$ in \cite{Chon:2021-12}. \cite{Vorobyov:2020-9} adopted a similar prescription as ours, but limited the cases with relatively high metallicity of $\geq 10^{-2}$\:$\Zsun$.
They show that the violent disc fragmentation continues longer at higher metallicities, by following the long-term evolution until the mass supply from the envelope to the disc ceases. Their results present a similar trend to ours at early phases of $\sim 10^4$\:yr after the disc formation. 
In \cite{Chon:2021-12}, at $\leq 10^{-3}$\:$\Zsun$ a large number of clumps form in $\sim10^3$\:yr after the protostar formation, similar to our results. 
Unlike our case, however, they observed a further increase in the clump number at relatively high metallicities of $\ge 10^{-2}$\:$\Zsun$.
This may be due to \cite{Chon:2021-12}'s omission of the stellar irradiation.
\cite{Bate:2014,Bate:2019} simulated the star cluster formation with a similar initial condition as in \cite{Chon:2021-12},
and reported that the stellar irradiation indeed suppresses low-mass star formation at $\gtrsim 10^{-2}$\:$\Zsun$.

%%%%%%%%%%%%%%%%%%%%%%%%%%%%%%%%%%%%%%%%%%%%
%%%%%%%%%%%%%%%%%%%%%%%%%%%%%%%%%%%%%%%%%%%%
%%% SECTION 5 %%%%
\section{Summary}
\label{Sec:5}

%------------------------------------------------------------------------------%

We have investigated the metallicity dependence of the circumstellar disc fragmentation by performing two-dimensional radiation-hydrodynamic simulations for a wide range of metallicities $1$--$0$\:$\Zsun$. 
Our simulations follow long-term evolution for 15\:kyr and in some models for 150\:kyr after disc formation by consistently solving the gas temperature from the energy equation with detailed thermal processes and chemistry network. In this manner, we have properly incorporated disc irradiation heating from an accreting protostar and examined its effect on the disc fragmentation. 
We have also separately solved gas and dust temperatures, which allows us to accurately follow the dust-induced fragmentation at very low metallicities. 
Our findings are summarized as follows.

%---------------------------------------------------------------------------%

\begin{itemize}
\item The disc fragmentation occurs in all the cases we examined by the end of simulations, although its frequency depends on metallicity (\Figsref{Fig:MassDis_1e-2}{Fig:Clump_1e-2}). 
The fragmentation is less frequent at $1$ and $10^{-1}$\:$\Zsun$, only providing a few clumps in a given moment. At lower metallicities of $\le 10^{-2}$\:$\Zsun$, we have observed intense fragmentation with $\gtrsim 10$ clumps at any time. In particular, the clump number is as high as $\simeq 20$ at $10^{-2}$--$10^{-5}$\:$\Zsun$. 

%--------------------------------------------------------------%

\item The clump-mass distribution depends strongly on the metallicity.
The ratio between the median clump mass and the central stellar mass is remarkably small for the cases of $10^{-2}$--$10^{-4}$\:$\Zsun$, where the number of the clumps is high. 
Those low-mass clumps with $\sim 10^{-2}$\:$\Msun$ are produced by the fragmentation of spiral arms in the disc induced by the dust cooling. 

%-----------------------------------------------------------------%

\item The stellar irradiation heating stabilizes the disc and reduces the clump number, especially at relatively high metallicities of $\gtrsim 10^{-2}$\:$\Zsun$. Even without stellar irradiation, the number of clumps tends to decrease with increasing metallicity $\gtrsim 10^{-2}$\:$\Zsun$. However, this trend is more pronounced by the irradiation heating, i.e. the irradiation heating strengthens the metallicity dependence of the disc fragmentation. 
\end{itemize}

%------------------------------------------------------------------%

In our simulations, the number of clumps is underestimated owing to limited spatial resolution, particularly in the cases of the lowest metallicities. At $10^{-3}$--$0$\:$\Zsun$, for instance, the number of clumps formed by disc fragmentation can be doubled by improving of the resolution. The actual metallicity dependence of the disc fragmentation should be stronger than in our results. This may explain the observed high close binary fraction of solar-type stars at low metallicities $-3<[{\rm Fe/H}]<0$.

%%%%%%%%%%%%%%%%%%%%%%%%%%%%%%%%%%%%%%%%%%%%%%%%%
%%%%%%%%%%%%%%%%%% ACKNOWLEDGMENTS %%%%%%%%%%%%%%%%%%

\section*{Acknowledgments}
The authors wish to express their cordial gratitude to Prof. Takahiro Tanaka, Leader of Innovative Area Grants-in-Aid for Scientific Research ``Gravitational wave physics and astronomy: Genesis'', for his continuous interest and encouragement.
The authors also would like to thank Profs. Masashi Chiba, Hidekazu Tanaka, and Kengo Tomida and Drs. Sunmyon Chon and Kazuyuki Sugimura for fruitful discussions and useful comments.
The numerical simulations were carried out on XC50  {\tt Aterui II} in Oshu City at the Center for Computational Astrophysics (CfCA) of the National Astronomical Observatory of Japan, through the courtesy of Prof. Eiichiro Kokubo.
This research could never be accomplished without the support by Grants-in-Aid for Scientific Research (KEIT: JP19K14760, JP19H05080, JP21H00058, JP21H01145, KO: 17H06360, 17H01102, 17H02869, 22H00149 TH:17H06360, 19H01934, 21H00041) from the Japan Society for the Promotion of Science. EIV acknowledges support from the Austrian Science Fund (FWF) under research grant P31635-N27. 
KO acknowledges support from the Amaldi Research
Center funded by the MIUR program "Dipartimento di
Eccellenza" (CUP:B81I18001170001).
We are deeply grateful to all parties involved.

%%%%%%%%%%%%%%%%%%%%%%%%%%%%%%%%%%%%%%%%%%%%%%%%%
%%%%%%%%%%%%%%%%%%% DATA AVAILABILITY% %%%%%%%%%%%%%%%%%%

\section*{Data availability}
The data underlying this article will be shared on reasonable request to the corresponding author.


%\bibliographystyle{mnras}
%\bibliography{references} % if your bibtex file is called example.bib

%%%%%%%%%%%%%%%%%%%%%%%%%%%%%%%%%%%%%%%%%%%%%%%%%%
%%%%%%%%%%%%%%%%%%%%%%%%%%%%%%%%%%%%%%%%%%%%%%%%%%

%%%%%%%%%%%%%%%%%%%%%%%%%%%%%%%%%%%%%%%%%%%%%%
%%%%%%%%%%%%%%%%% APPENDICES %%%%%%%%%%%%%%%%%%%%%
\appendix

\section{thermal evolution}
\label{App:thermal}

%------------------------------------------------------------------%

In \Secref{Sec:3-1}, we discuss the thermal evolution of the gas, 
especially focusing only on the circumstellar disc and its surroundings. 
In this section, we describe the thermal evolution of the cloud as a whole and which cooling process is important in each metallicity.

%%%%% FIGURE A1 %%%%%
\begin{figure*}
 \begin{center}
 \begin{tabular}{c} 
  {\includegraphics[width=1.95\columnwidth]{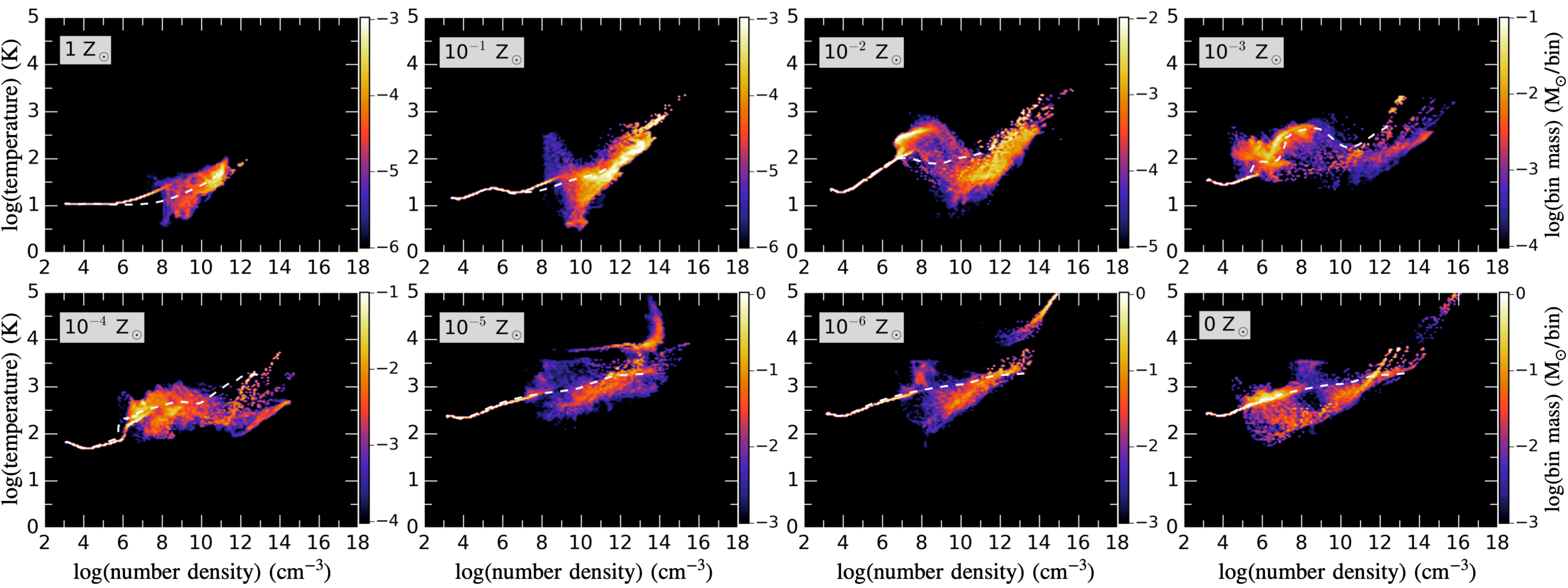}}
 \end{tabular}
 \caption{
 Gas mass distributions on the density-temperature phase diagrams at 15\:kyr elapsed from the disc formation. 
 The panels depict models with seven different metallicities shown in the upper left of each panel. 
 The colour indicates the mass in each density-temperature bin 
 with the widths of $\Delta \log (n_{\mathrm{H}}/{\rm cm^{-3}}) = 0.1$ and $\Delta \log (T/{\rm K}) = 0.03$. 
 The white dashed line is the density-temperature distribution at the disc formation. 
 }
 \label{Fig:nT_massgrid}
 \end{center}
\end{figure*}
%

%------------------------------------------------------------------%

\Figref{Fig:nT_massgrid} shows the gas mass distribution on the density-temperature phase diagram. 
The trend toward higher temperatures for lower metallicities is observed in \Figref{Fig:Temperature_1e-2}, 
and it is the same in \Figref{Fig:nT_massgrid}. 
The gas has the temperature of $10$--$100$\:K at $1$\:$\Zsun$, 
while it has the high temperature of $100$--$10^5$\:K at $0$\:$\Zsun$. 
This difference in temperature range is due to the variance in the cooling process as explained in \Secref{Sec:3-1}. 

%------------------------------------------------------------------%

At $\ge 10^{-1}$\:$\Zsun$, 
the temperature is $\sim$10\:K at the density of $10^3$--$10^6$\:cm$^{-3}$ due to the efficient cooling by metal line and dust emission. 
The temperature increases at higher densities of $\gtrsim 10^6$\:cm$^{-3}$ due to the stellar irradiation and the compressional heating (see also \Secref{Sec:4-1}). 

%------------------------------------------------------------------%

At $\le 10^{-2}$\:$\Zsun$, the cooling by metal line and dust emission becomes inefficient, 
resulting in molecular-line emissions as the main cooling process. 
In this case, the temperature is $30$--$300$\:K in the low-density region of $10^3$--$10^6$\:cm$^{-3}$. 
In addition, at $10^{-2}$--$10^{-4}$\:$\Zsun$, the temperature rises sharply to $\sim 500$\:K at the density of $10^6$--$10^7$\:cm$^{-3}$ 
because of the chemical heating associated with the H$_2$ formation. 
Then, the temperature peaks at the density of $\sim 10^8$\:cm$^{-3}$. 
At higher densities, the temperature decreases as the dust cooling begins to take effect. 
This temperature decrease, which is also seen in \Figref{Fig:nT_1e-2}, determines the typical clump mass of $10^{-3}$\:$\Zsun$.

%------------------------------------------------------------------%

At even lower metallicities of $\le 10^{-6}$\:$\Zsun$, 
the dust cooling does not work, 
and the gas is cooled by H$_2$-line emission in the whole density of $10^3$--$10^{14}$\:cm$^{-3}$. 
Since the H$_2$-line cooling is less efficient than the dust cooling, 
the temperature is high, at $\sim10^3$\:K. 
The hot distribution of $>10^4$\:K in the denser region of $>10^{14}$\:cm$^{-3}$
represents the second object seen in \Figref{Fig:Surface_binary},
where the cooling is artificially suppressed due to the resolution limit (Equation \ref{Eq:Climit}).

%%%%%%%%%%%%%%%%%%%%%%%%%%%%%%%%%%%%%%%%%%%%
%%% Appendix B %%%%
\section{Identifying self-gravitating clumps}
\label{App:frag}

Here, we describe the method for identifying clumps from calculation results. 
We first search for a grid that has a higher surface density than the eight neighbouring grids in the computational domain. 
We call such a grid the peak grid.
The Jeans length $\lambda_{\rm J}$ at the peak grid is 
\begin{align}
\lambda_{\rm J} = 2\frac{c_{\rm s,peak}^2}{G \Sigma_{\rm peak}},
\end{align}
where $c_{\rm s,peak}$ and $\Sigma_{\rm peak}$ are the sound velocity and the surface density at the peak grid. 
We note that this definition of the Jeans length $\lambda_{\rm J}$ is consistent with another definition of $l_{\rm J}$ in \Eqref{Eq:Truelove} when the gas self-gravity becomes significant.
The Jeans radius $R_{\rm J}$ is half of the Jeans length, $R_{\rm}=\lambda_{\rm J}/2$. 
To ensure that the peak density $\Sigma_{\rm peak}$ is high enough to qualify as a ``clump'', we check the following condition,
\begin{align}
(\rm{i})~\Sigma_{\rm peak} \ge 10\Sigma_{\rm avr},
\end{align}
where $\Sigma_{\rm avr}$ is the average surface density of the annulus between the radii of $R_{\rm J}$ and $1.5R_{\rm J}$ from the peak grid.
If this condition (i) is satisfied, 
an aggregate of the grids within the Jeans radius of the peak grid is a clump candidate. 

%------------------------------------------------------------------%

Next, we check if the clump candidate is gravitationally bound. 
Based on the virial analysis,
we set the gravitationally-bound condition as
\begin{align}
({\rm ii})~\frac{|E_{\rm g}|}{2(\gamma-1)E_{\rm th}+2E_{\rm kin}} \ge 0.6,
\label{Eq:condition_ii}
\end{align}
where $E_{\rm g}$ is the gravitational energy, $E_{\rm th}$ is the thermal energy, and $E_{\rm kin}$ is the kinetic energy
of the clump candidate. 
We explored the values on the right-hand side of \Eqref{Eq:condition_ii} between $0.1$ and $1.0$.
Then, we selected the value of $0.6$ that can be judged without missing the candidate that is a visually obvious clump. 
The three energies are given by 
\begin{align}
&E_{\rm g}   = - \int\int\frac{1}{2}G\frac{\Sigma \Sigma'}{|\bm{r}-\bm{r}'|}{\rm d}S {\rm d}S', \\
&E_{\rm th}  = \int e {\rm d}S , \\
&E_{\rm kin} = \int \frac{1}{2}\Sigma\left( \bm{u}-\bm{u}_{\rm c} \right)^{2} {\rm d}S, 
\end{align}
where $\bm{u}_{\rm c}$ is the velocity of the mass centre of the clump candidate. 
The integral range is within the Jeans radius $R_{\rm J}$. 
We declare the clump candidate to be an actual clump if it satisfies the condition (ii).

%-----------------------------------------------------------------------%

Some clumps are ejected from the central region due to gravitational interactions 
with other clumps as well as the central star.
Such an ejected clump moves toward the outer region, where the grid is coarser in the spherical coordinate.
With insufficient resolution, the density of the ejected clump artificially decreases, reducing its gravitational energy.
Consequently, the ejected clump no longer satisfies the gravitationally-bound condition (ii).
However, we expect such ejected clumps to keep sufficiently high density in reality.
Therefore, we added the following conditions to the clump identification to avoid the omission of counting ejected clumps.
Among the candidates satisfying the density condition (i) but not the condition (ii),
we certify one as a clump if the speed of the mass centre $u_{\rm c}$ exceeds the escape velocity from the central stellar gravity $u_{\rm esc}$, i.e. 
\begin{align}
({\rm iii})~u_{\rm c} > u_{\rm esc} = \sqrt{2\frac{GM_{\ast}}{r_{\rm c}}},
\end{align}
where $r_{\rm c}$ is the distance between the central star and the mass centre of the candidate.
At $t=15$\:kyr from the disc formation (\Figref{Fig:MassDis_1e-2}),
we identified the ejected clumps satisfying the condition (iii) at $10^{-2}$, $10^{-3}$, $10^{-6}$, and $0$\:$\Zsun$, whose ratios are accounted for $\sim 20$--$40$ \% of the total clump number.

%%%%%%%%%%%%%%%%%%%%%%%%%%%%%%%%%%%%%%%%%%%%
%%% Appendix C %%%%
\section{Effects of varying initial properties of star-forming clouds}
\label{App:parameter}

%%%%% FIGURE C1 %%%%%
\begin{figure*}
 \begin{center}
 \begin{tabular}{c} 
  {\includegraphics[width=1.95\columnwidth]{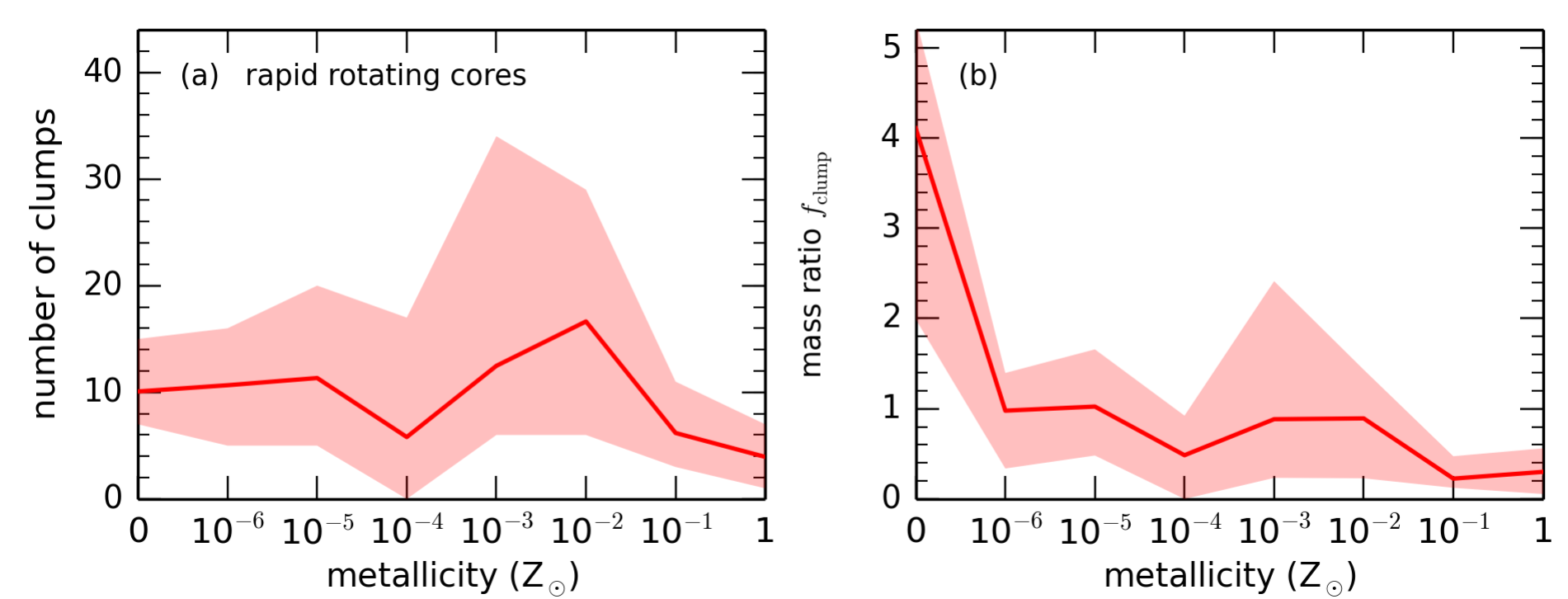}}
 \end{tabular}
 \caption{Same as \Figref{Fig:Clump_1e-2} but for the cases starting from the rapidly rotating cores.
 }
 \label{Fig:Clump_5e-2}
 \end{center}
\end{figure*}

We here briefly describe the disc fragmentation in additional simulations starting from different initial cloud core models: rapidly rotating cores (\Appref{App:Rapid}) and massive cores (\Appref{App:Massive}). As in \Secref{Sec:3}, we pay special attention to the number of clumps and the mass ratio $f_{\rm clump}$.

%%%%%%%%%%%%%%%%%%%%%%%%%%%%%%%%%%%%%%%%%%%%
%%% Appendix C-1 %%%%
\subsection{rapidly rotating cores}
\label{App:Rapid}

We investigate the cases of ``rapidly rotating cores'', for which we assume a higher initial spin than the standard cases described in \Secref{Sec:2-2}.
While we have assumed the ratio of the rotational and gravitational energies of $\beta = 0.01$ for the standard case, we here consider the cases with $\beta = 0.05$ below. These values are comparable to those assumed in previous studies, $\beta = 0.01$--$0.02$ in \cite{Chiaki:2022}, and 0.03--0.09 in \cite{Shima:2021}. The other settings are identical to those used for the standard cases with different metallicities. 
We follow the evolution for 15\:kyr after the disc formation for all the cases.

%----------------------------------------------------------------------------%

\Figref{Fig:Clump_5e-2}a shows the variations of the clump number with different metallicities for these cases (cf. \Figref{Fig:Clump_1e-2}a). We see that the number of clumps is largest at $10^{-2}$--$10^{-3}$\:$\Zsun$ and it decreases toward both the high- and low-metallicity ends. While the time-averaged clump number has a peak value of $\simeq 20$ at $10^{-2}$\:$\Zsun$, it decreases to a few in the high-metallicity tail of $10^{-1}$--$1$\:$\Zsun$. The overall trends are similar to the metallicity dependence of the standard cases (\Secref{Sec:3-1}).

%------------------------------------------------------------%

\Figref{Fig:Clump_5e-2}b shows the metallicity dependence of the mass ratio $f_{\mathrm{clump}}$, the ratio of the total clump mass to the central stellar mass (cf. \Figref{Fig:Clump_1e-2}b). The mass ratio $f_{\mathrm{clump}}$ takes the smallest values of $\simeq 0.3$ at high metallicities $1$ and $10^{-1}$\:$\Zsun$, and it is nearly constant at 0.5--1.0 for the cases of $10^{-2}$--$10^{-6}$\:$\Zsun$. Only the value at $0$\:$\Zsun$ is remarkably larger than those for the other cases, indicating the formation of a massive secondary clump orbiting around the central star (\Figref{Fig:Surface_binary}). Again, the overall trend is similar to that for the standard cases (\Secref{Sec:3-1}). 

%------------------------------------------------------------%

%%%%% FIGURE C2 %%%%%
\begin{figure*}
 \begin{center}
 \begin{tabular}{c} 
  {\includegraphics[width=1.95\columnwidth]{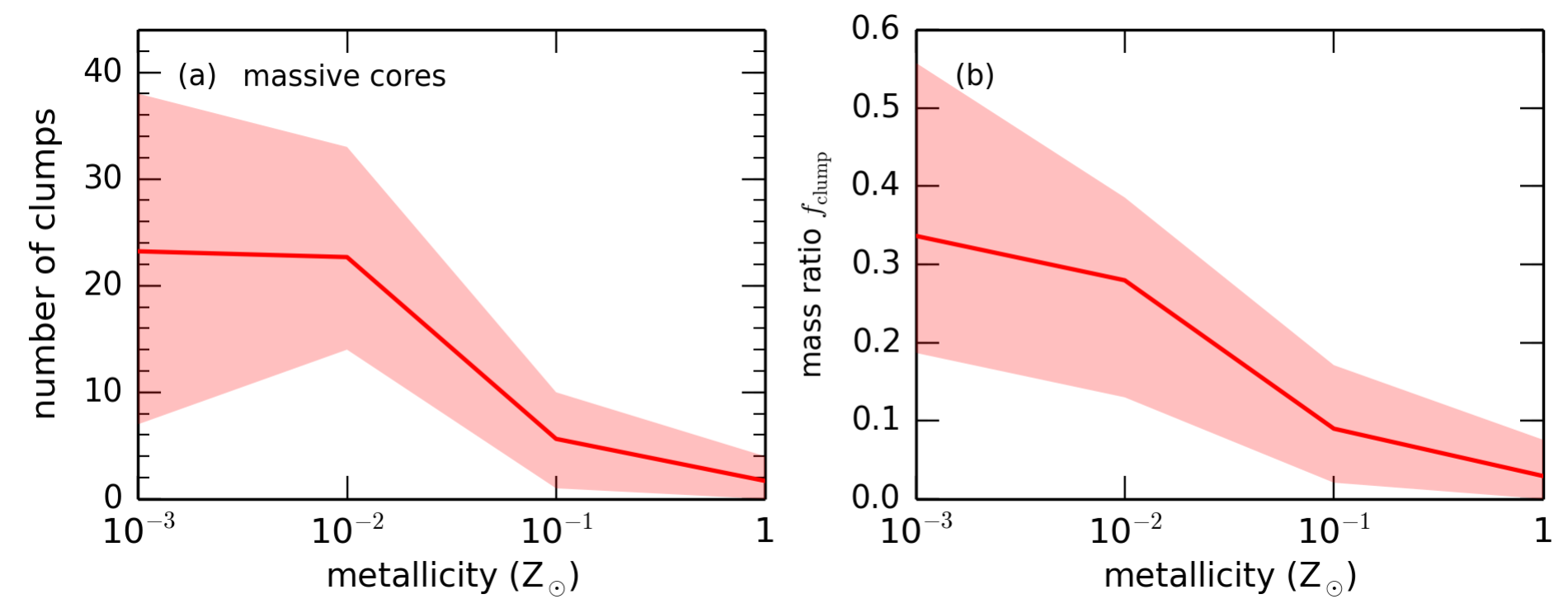}}
 \end{tabular}
 \caption{
 Same as \Figref{Fig:Clump_1e-2} but for the cases starting from the initial massive cores. 
 }
 \label{Fig:Clump_massive}
 \end{center}
\end{figure*}
%

%------------------------------------------------------------%

It is not straightforward to extract the effects of varying $\beta$ on the disc fragmentation by comparing \Figref{Fig:Clump_1e-2} and \Figref{Fig:Clump_5e-2} at each metallicity.
The higher $\beta$ results in a larger disc size at a given epoch, leading to effectively poorer spatial resolution in the outer disc, where fragmentation often occurs.
Since our simulations tend to underestimate the number of clumps with the lower resolution (\Secref{Sec:4-2}), it is tricky to draw out the effect of the initial spin, considering the effect of the different spatial resolutions. We, nonetheless, find no clear and qualitative effects of varying the initial spin on the metallicity dependence of the disc fragmentation.

%%%%%%%%%%%%%%%%%%%%%%%%%%%%%%%%%%%%%%%%%%%%
%%% Appendix C-2 %%%%
\subsection{massive cores}
\label{App:Massive}

%------------------------------------------------%

For our standard cases considered in \Secref{Sec:2-2}, we have assumed the less massive initial cores for the higher metallicities, considering the Jeans mass scale given by the gas thermal evolution at each metallicity. In reality, however, the star formation may rarely start with more massive cloud cores than the standard cases, for instance, massive cores supposed for the present-day massive star formation \citep[][]{McKee:2003}. We here perform additional simulations for such rare cases, using the same initial core of $10^{-3}$\:$\Zsun$ with $\simeq 60$\:$\Msun$ (Table \ref{Tab:2_LM}) for higher-metallicity cases of $1$, $10^{-1}$, and $10^{-2}$\:$\Zsun$. The initial mass is $\sim 20$ times larger than that for the standard case of $1$\:$\Zsun$, $\simeq 3.4$\:$\Msun$ (Table \ref{Tab:2_LM}). We do not consider the cases with $< 10^{-3}$\:$\Zsun$, because the initial core mass becomes extremely massive at  $100$--$1000$\:$\Msun$, corresponding to cluster-forming clumps at $1$\:$\Zsun$.

%----------------------------------------------------------------------------%

\Figref{Fig:Clump_massive} shows the metallicity dependencies of the clump number and the mass ratio $f_{\mathrm{clump}}$ for these ``massive core'' cases (cf. \Figref{Fig:Clump_1e-2}). The time-averaged number of clumps tends to increase with decreasing the metallicity, from a few at $1$-$10^{-1}$\:$\Zsun$ to $>20$ at $10^{-2} - 10^{-3}$\:$\Zsun$ (\Figref{Fig:Clump_massive}a). 
The mass ratio $f_{\mathrm{clump}}$ also increases with decreasing the metallicity (\Figref{Fig:Clump_massive}b). 
These trends are almost identical to those in \Figref{Fig:Clump_1e-2}. 

%------------------------------------------------------------%

Note that, among the standard cases, the mass-supply rate from the envelope to the disc is 
significantly lower at the higher metallicities (\Secref{Sec:3}).
In fact, in \Figref{Fig:accretion_rate}, 
the accretion rate to the central star at $1$\:$\Zsun$ is about 10 times smaller than that at $10^{-3}$\:$\Zsun$. 
On the other hand, the accretion rates at $1$ and $10^{-3}$\:$\Zsun$ in the cases of massive cores are almost the same. 
Interestingly, even though the accretion rates are not much different among the massive cores,
\Figref{Fig:Clump_massive} shows the same metallicity dependence as in \Figref{Fig:Clump_1e-2}. 
\cite{Tanaka:2014} analyzed the disc stability using the one-dimensional steady accretion model, 
and showed that the disc becomes more stable at higher metallicities in the range of $1$--$10^{-3}$\:$\Zsun$, regardless of the prestellar cloud mass.
Our results support their prediction.

%%%%%%%%%%%%%%%%%%%%%%%%%%%%%%%%%%%%%%%%%%%%%%%%%%

% Don't change these lines
\bsp	% typesetting comment
\label{lastpage}
\end{document}